\documentclass[superscriptaddress,nobibnotes,amsmath,amssymb,notitlepage,twocolumn,prb,longbibliography]{revtex4-2}

\usepackage{bm,braket}
\usepackage[toc,page]{appendix}
\usepackage{comment}
\usepackage{dcolumn}

\usepackage{amsfonts}

\usepackage{graphicx,color,hyperref}
\usepackage[caption=false]{subfig}
\hypersetup{colorlinks=true, linkcolor=blue, citecolor=blue, urlcolor=blue} 

\graphicspath{{./img2/}}

\usepackage{amsmath}
\usepackage{amssymb}
\usepackage{bbm}
\usepackage{physics}
\usepackage{float}
\usepackage{graphicx}
\usepackage{dcolumn} 
\usepackage{bm} 
\usepackage{siunitx}
\usepackage{enumitem}  
\usepackage{mathtools}
\usepackage[pdftex,outline]{contour}
\contourlength{1pt}

\newcommand{\beq}[1]{\begin{equation}\label{#1}}
\newcommand{\eep}{\;.\end{equation}}
\newcommand{\eec}{\;,\end{equation}}
\newcommand{\eeq}{\end{equation}}

\renewcommand{\k}{\kappa}

\newcommand{\om}{\omega}

\DeclareMathAlphabet{\mathcal}{OMS}{cmsy}{m}{n} 

\renewcommand{\vec}[1]{{\vb #1}}

\newcommand{\kv}{\vec{k}}
\newcommand{\R}{\vec{R}}
\newcommand{\rv}{\vec{r}}

\DeclareMathOperator{\Pf}{Pf}

\usepackage{amsmath}
\usepackage{amssymb}
\usepackage{xcolor}
\usepackage{bbm}
\usepackage{physics}
\usepackage{float}
\usepackage{graphicx}
\usepackage{dcolumn} 
\usepackage{bm} 
\usepackage{siunitx}
\usepackage{enumitem}  

\makeatletter
\renewcommand*{\fnum@figure}{{\normalfont\bfseries \figurename~\thefigure}}
\makeatother

\allowdisplaybreaks

\definecolor{orange}{rgb}{1,0.5,0}

\newcommand{\sect}[1]{\vspace{0.3em}{\it #1.}---}

\graphicspath{{./img2/}}

\DeclareMathAlphabet{\mathcal}{OMS}{cmsy}{m}{n} 

\newcommand{\intBZ}{\int_{\text{BZ}}} 

\newcommand{\ii}{\mathrm{i}}
\newcommand{\ee}{\mathrm{e}}

\newcommand{\jj}{\mathrm{j}}
\newcommand{\kk}{\mathrm{k}}
\newcommand{\KK}{\boldsymbol{\kappa}}

\renewcommand{\aa}{\mathrm{a}}
\newcommand{\Eu}{\mathrm{Eu}}
\DeclareMathOperator{\diag}{diag}
\DeclareMathOperator{\sgn}{sgn}
\newcommand{\hh}{\mathcal{H}}
\newcommand{\hhc}{\mathcal{H}_{\mathbb{C}}}
\renewcommand{\dh}{\hat{\vb{d}}}
\newcommand{\nh}{\hat{\vb{n}}}

\newcommand{\BZ}{\text{BZ}}

\makeatletter
\newcommand{\specificthanks}[1]{\@fnsymbol{#1}}
\makeatother

\begin{document}

\preprint{APS/123-QED}

\title{Non-Abelian Hopf-Euler insulators}

\author{Wojciech J. Jankowski}
\email{wjj25@cam.ac.uk}
\thanks{These authors contributed equally.}
\affiliation{TCM Group, Cavendish Laboratory, Department of Physics, J J Thomson Avenue, Cambridge CB3 0HE, United Kingdom\looseness-1}

\author{Arthur S. Morris}
\email{asm99@cam.ac.uk}
\thanks{These authors contributed equally.}
\affiliation{TCM Group, Cavendish Laboratory, Department of Physics, J J Thomson Avenue, Cambridge CB3 0HE, United Kingdom\looseness-1}

\author{Zory Davoyan}
\thanks{}
\affiliation{TCM Group, Cavendish Laboratory, Department of Physics, J J Thomson Avenue, Cambridge CB3 0HE, United Kingdom\looseness-1}

\author{Adrien Bouhon}
\thanks{}
\affiliation{TCM Group, Cavendish Laboratory, Department of Physics, J J Thomson Avenue, Cambridge CB3 0HE, United Kingdom\looseness-1}
\affiliation{Nordita, Stockholm University and KTH Royal Institute of Technology, Hannes Alfv{\'e}ns v{\"a}g 12, SE-106 91 Stockholm, Sweden}

 \author{F. Nur \"Unal}
\thanks{}
\affiliation{TCM Group, Cavendish Laboratory, Department of Physics, J J Thomson Avenue, Cambridge CB3 0HE, United Kingdom\looseness-1}

\author{Robert-Jan Slager}
\email{rjs269@cam.ac.uk}
\affiliation{TCM Group, Cavendish Laboratory, Department of Physics, J J Thomson Avenue, Cambridge CB3 0HE, United Kingdom\looseness-1}

\date{\today}

\begin{abstract}
  We discuss a class of three-band non-Abelian topological insulators in three dimensions that carry a single bulk Hopf index protected by spatiotemporal ($\mathcal{PT}$) inversion symmetry. These phases may also host subdimensional topological invariants given by the Euler characteristic class, resulting in real Hopf-Euler insulators. Such systems  naturally realize helical nodal structures in the three-dimensional Brillouin zone, providing a physical manifestation of the linking number described by the Hopf invariant. We show that, by opening a gap between the valence bands of these systems, one finds a fully-gapped ``flag'' phase, which displays a three-band multi-gap Pontryagin invariant. Unlike the previously reported $\mathcal{PT}$-symmetric four-band real Hopf insulator, which hosts a $\mathbb{Z} \oplus \mathbb{Z}$ invariant, these phases are not unitarily equivalent to two copies of a complex two-band Hopf insulator. We show that such uncharted phases can be obtained through dimensional extension of two-dimensional Euler insulators, and that they support (i) an optical bulk integrated circular shift effect quantized by the Hopf invariant, (ii) quantum-geometric breathing in the real space Wannier functions, and (iii) surface Euler topology on boundaries.~Consequently, our findings pave the way for novel experimental realizations of real-space quantum-geometry, as these systems may be directly simulated by utilizing synthetic dimensions in metamaterials or ultracold atoms. 
\end{abstract} 

\maketitle

\section{Introduction}\label{sec::I}
Non-Abelian phenomena and gauge structures are of broad interest in contexts ranging from condensed matter to high energy physics. Such non-commuting objects can induce a wide range of complex phenomena, many of which have no Abelian counterpart. A salient example is provided by non-Abelian anyons, which can exhibit exotic braiding statistics; these are, moreover, an active area of research due to their potential application in  topological quantum computation ~\cite{RevModPhys.80.1083}.

It is not only quantum-mechanical systems that can realize non-Abelian gauge fields. For instance, classical soft matter systems can host non-Abelian topological defects in the form of $\pi$-disclinations within biaxial nematic liquid crystals~\cite{volovik2018investigation, PhysRevX.6.041025, Beekman20171, Kamienrmp}. This example is of particular importance due to the recent discovery that band degeneracies in systems possessing spatiotemporal inversion ($\mathcal{PT}$) or $\mathcal{C}_2\mathcal{T}$ (two-fold rotation with time-reversal) symmetry~\cite{doi:10.1126/science.aau8740, Bouhon_2019,PhysRevX.9.021013,  ahnprl, bouhon2018wilson} can host non-Abelian charges in a fashion directly analogous to the emergence of $\pi$-disclination vortices in biaxial nematics. In this scenario, the charges are defined by the particular type of rotation exhibited by the Bloch eigenstates $\ket{u^a(\vb{k})}$ in the vicinity of the nodes, differing relative to each other. Furthermore, these band degeneracies may be braided around each other to produce band subspaces (groups of bands) that host similarly charged nodes that cannot be mutually annihilated. Such processes hence produce a novel multi-gap phase~\cite{Bouhon2020_geo} in which the two-band subspace exhibits a multi-gap topological invariant, the Euler class~\cite{Bouhon_2019,PhysRevX.9.021013}. These multi-gap phases in principle go beyond conventional single-gap topological phases~\cite{Rmp1,Rmp2, clas1}, which can be classified by comparing how irreducible band representations glue together over the Brillouin zone (BZ)~\cite{fukane, Slager_NatPhys_2013, Kruthoff_2017, rjs_translational} and comparing their real space Wannier description~\cite{Po2017, Bradlyn:2017}, as they are in general not symmetry-indicated~\cite{Bouhon2020_geo}. Notably, multi-gap invariants, such as the Euler class $\chi$, and the corresponding braiding of band degeneracies in two-dimensional systems have been related to a variety of physical systems and phenomena, including out-of-equilibrium quenches and Floquet systems~\cite{Unal_2020,slager2024floquet,breach2024interferometry}, phonon modes~\cite{Peng2021, Peng2022Multi}, magnetic systems~\cite{mag2024euler,magnetic}, and implementations in metamaterials~\cite{Jiang_2021, natcom4band,JIANG2024, Guo1Dexp}.

In three spatial dimensions, an assortment of different multi-gap phases are possible~\cite{Bouhon2020_geo, bouhon2023quantum, bouhon2022multigap}, all of which lie outside the paradigm of K-theory and single-gap stable equivalence~\cite{Kitaevtenfold, Kruthoff_2017}; these phases are instead described by homotopy groups, which capture the fine topological detail of few-band systems~\cite{Bouhon2020_geo}. In this case, the type of topology that can be realized depends sensitively upon how the bands are partitioned~\cite{Bouhon2020_geo}, since these subspaces determine the classifying space into which the Bloch Hamiltonian defines a map. The particular topological class to which a given system belongs may be determined by computing the corresponding charges induced on this manifold by its Hamiltonian~\cite{Bouhon2020_geo,Chiu_2016}. For example, in a four-band system at three-quarters filling one may compute the Pontryagin index, which also characterizes non-Abelian SU(2) instantons in Yang-Mills theory~\cite{PhysRevB.109.165125}. If, in addition, all occupied and unoccupied bands in this system are initially mutually well-separated in energy, then, upon the introduction of band crossings, it is possible to assign the system a homotopy charge in the group ${\pi_3[S^3] \cong \mathbb{Z}}$, and moreover to realize a braiding protocol with split-biquaternionic charges~\cite{PhysRevB.109.165125}.

In the same vein, Hopf insulators, which provide a solid-state realization of the Hopf fibration ${S^1\xhookrightarrow {} S^3\overset{\pi}{\rightarrow} S^2}$ \cite{Hopf_1, Hopf_2,Hopf_3, Kennedy_2016, Unal2019, alexandradinata2021,Nelson_2022, Zhu_2023, Lim2023}, also fall beyond the stable equivalence classification captured by K-theory. The two-band Hopf insulator phase is characterized by a single integer-valued topological quantum number, namely the Hopf invariant $\hhc$~\footnote{To avoid confusion between the Euler class and the Hopf invariant, which are both usually written as $\chi$, we have chosen to use the symbol $\hh$ to denote the latter quantity.}, which takes values in the homotopy group $\pi_3[S^2] \cong \mathbb{Z}$. In this context, the two sphere is the classifying space of a two-band complex system at half-filling, 
\begin{equation}
    \mathsf{Gr}_{1, 2}(\mathbb{C})\cong \frac{\mathsf{U}(2)}{\mathsf{U}(1)\times\mathsf{U}(1)}\cong S^2.
\end{equation}
The Hopf invariant of this model may be computed from the following integral over the three-dimensional Brillouin zone, $\BZ\cong T^3$:
\begin{equation}\label{eq:Hopf}
   \hhc = -\frac{1}{4\pi^2} \intBZ A\wedge F,
\end{equation}
where $A = \ii\braket{u}{\dd u}$ is the Abelian Berry connection of the occupied band $\ket{u(\vb{k})}$, and $F = \dd A$ is the corresponding curvature. While the original study on the Hopf insulator considered only this bulk index, subsequent work has investigated the consequences of the presence of additional ``weak'' invariants on the two-dimensional coordinate planes within the 3D BZ. It turns out that, when the Chern numbers on the $k_x$-, $k_y$-, and $k_z$-planes are $\vb{C} = (C_x, C_y, C_z)$, respectively, the Hopf invariant is instead an element of the set $\mathbb{Z}_{2\gcd(\vb{C})}$, where $\gcd(\vb{C})=\gcd(C_x, C_y, C_z)$ is the greatest common divisor of the integers $C_x$, $C_y$, and $C_z$~\cite{Kennedy_2016}.

Similarly to other topological invariants, the presence of a non-trivial Hopf invariant in the bulk of a system has consequences for its response functions. In particular, in the presence of a static electromagnetic field, the vacuum of a three-dimensional Hopf insulator may support a topological magnetoelectric effect~\cite{Qi_2008, alexandradinata2021}. In general, this phenomenon is described by the effective action for axion electrodynamics~\cite{Qi_2008, axion1, Axion2, Axion3,Axion4}
\beq{eq:axion}
    S_{\text{axion}} = \frac{\theta}{16\pi^3} \int \mathcal{F} \wedge \mathcal{F},
\eeq
where $\mathcal{F}$ is the electromagnetic Maxwell tensor. Here the so-called ``$\theta$-angle'', which is a property of the medium, can be obtained from the integral of the Chern-Simons form over the BZ
\begin{equation}\label{eq::CS-kspace}
    \theta = \frac{1}{4\pi^2} \intBZ  \Tr[A\wedge\dd A+\frac{2}{3} A\wedge A\wedge A] \phantom{n}(\text{mod}\, 2\pi), 
\end{equation}
where $A_{ab}=\ii\braket{u_a}{\dd u_b}$ is the non-Abelian Berry connection and the trace is evaluated over the band indices $a,b$. In the context of the two-band Hopf insulator, Eq.~\eqref{eq::CS-kspace} reduces to (a multiple of) Eq.~\eqref{eq:Hopf}, and we see that ${\theta = \pi \hhc\,(\text{mod}\,2\pi)}$. It should be stressed that only the ground states with $\theta = \pi~(\text{mod}~2\pi)$ display a topological magnetoelectric effect~\cite{alexandradinata2021}; this follows from the variation of the action $S_\text{axion}$ in Eq.~\eqref{eq:axion} under large gauge transformations, which change $\theta \rightarrow \theta + 2\pi$. Nonetheless, non-trivial quantized optical responses can emerge even in magnetoelectrically trivial media, a concrete example being the $\mathcal{PT}$-symmetric Hopf insulator with $\theta = 0~(\text{mod}~2\pi)$~\cite{jankowski2024quantized}.

The Hopf map also arises in a number of other topological phases. For instance, it has strong connections to ultracold atoms, where it has been shown to arise in quenched Chern bands \cite{Tarnowski_2019} and Euler systems \cite{Unal_2020, Zhao_2022}. One generalization is the $N$-band complex Hopf insulator~\cite{lapierre2021}, in which a Hopf invariant may be assigned to an isolated two-band subspace which is separated from the rest of the space by gaps both above and below it.~Of particular relevance to the present work is the four-band real Hopf insulator (RHI), introduced in Ref.~\cite{Lim2023}, which is realized in half-filled systems satisfying a reality condition. In such systems, the (oriented cover of the) classifying space is isomorphic to a pair of spheres~\cite{Bouhon2020_geo, Lim2023, bouhon2023quantum},
\begin{equation}
    \mathsf{Gr}^+_{2, 4}(\mathbb{R}) \cong \frac{\mathsf{SO}(4)}{\mathsf{SO}(2)\times\mathsf{SO}(2)}\cong S^2_-\times S^2_+,
\end{equation}
which gives rise to two intertwined Hopf invariants, $\hh_{\pm}$. Furthermore, the integrated shift photoconductivities of these systems, which characterize their coupling to circularly polarized light, have recently been shown to be quantized~\cite{jankowski2024quantized}.

In this work we establish further results concerning real multi-band topological phases, in particular Hopf-Euler phases, in more general settings. We proceed by introducing three-band real Hopf insulators, and moreover discuss the three- and four-band real phases which carry a Hopf index in the presence of a non-trivial Euler class on one or more of the coordinate planes within the BZ. We also identify and discuss particular physical manifestations of such distinct phases, namely: (\textit{i}) a bulk quantized non-linear optical circular shift response; (\textit{ii}) real-space oscillations of maximally-localized bulk Wannier functions, (\textit{iii}) boundary states hosting Euler topology at the surface, and (\textit{iv}) nodal helices naturally realized in the presence of the non-trivial weak Euler invariants.

The manuscript is organized as follows. In Sec.~\ref{sec::II}, we introduce a set of distinct homotopy-classified non-Abelian $\mathcal{PT}$-symmetric real Hopf phases in three spatial dimensions, with three rather than four bands~\cite{Lim2023}. In this context, we additionally identify fully-gapped `flag' phases which possess a strong homotopy invariant associated with all three bands, which may be identified with a Hopf index. Finally, as a more central component of this work, we introduce Hopf-Euler insulators, which host the aforementioned Euler class invariants on two dimensional sections of BZ, in addition to non-trivial Hopf topology. In Sec.~\ref{sec::III}, we identify a manifestation of the interplay between the strong Hopf and weak Euler invariants which is realized in these phases, namely the presence of nodal helices. The existence of these nodal structures is a natural consequence of a fundamental mathematical connection between the topological invariants and the preimages of the maps which characterize the Hamiltonians. Furthermore, in Sec.~\ref{sec::III}, we examine the physical phenomena displayed by these non-Abelian phases. We show that the bulk Hopf index is reflected in the non-linear optical response of the system (specifically the quantized integrated shift response), and that the quantum geometry of these phases emerges in the form of a quantum-geometric breathing (QGB) of maximally-localized hybrid Wannier functions in real space, that is, oscillations of their second moments. Lastly, we show that the bulk real Hopf invariants induce the Euler class in the boundary states by means of the teleportation of Euler curvature, provided the boundary preserves the $\mathcal{C}_2\mathcal{T}$ symmetry which is respected in the bulk. In Sec.~\ref{sec::IV}, we provide concrete realizations of the introduced three- and four-band Hopf/Euler phases in minimal models. We elaborate on possible experimental realizations of the aforementioned phases in Sec.~\ref{sec::V}. Finally, we discuss our results in Sec.~\ref{sec::VI}, where we review the connections between the homotopy-classified two-band, three-band, and four-band phases which arise from dimensional extensions and complexification relations.~We then examine electromagnetic responses in multi-gap phases, before concluding in Sec.~\ref{sec::VII}.  

\section{Hopf-Euler phases}\label{sec::II} 
In this section we utilize the Pontryagin-Thom construction to classify three- and four-band $\mathcal{PT}$-symmetric gapped phases of matter with Hopf indices in three dimensions. We first examine three-band phases with a single gap, and we demonstrate that such phases are classified by a Hopf invariant and three Euler classes. Of particular note are the Hopf-Euler phases, in which both of these topological invariants are simultaneously non-trivial. We then explore the modifications that appear when imposing the additional condition that the occupied bands be gapped from one another, before concluding with a discussion of the extension of the three-band phases to four-band systems.

\subsection{Three-band Hopf-Euler phases}\label{sec:3BandPhases}
Let us begin by describing the topological invariants that may be assigned to a real three-band model in three dimensions. Let $H_3(\vb{k})$ be a $3\times 3$ real Bloch Hamiltonian, where $\vb{k}=(k_x, k_y, k_z) \in \BZ\cong T^3$ is the quasi-momentum which takes values in the Brillouin zone, a 3-torus. The reality of the Hamiltonian may be ensured by the presence of particular symmetries, for example, $\mathcal{PT}$ symmetry. We will choose a gauge in which the Bloch Hamiltonian is a periodic function over the BZ, so that $H_3(\vb{k}+\vb{G})=H_3(\vb{k})$ for any reciprocal-lattice vector $\vb{G}$. We denote the eigenvectors of $H_3(\vb{k})$ as $\ket{u_a(\vb{k})}$, $a=1, 2, 3$, while the corresponding energies are $E_a(\vb{k})$. Since the eigenvectors $\ket{u_a}$ may be chosen to be real, we sometimes refer to both $\ket{u_a}$ and its dual simply as $\vb{u}_a$.~For it to be possible to ascribe a topological class to the system, it must be gapped, and correspondingly we assume that $E_3(\vb{k})> E_{1, 2}(\vb{k})$ for all $\vb{k}$. Moreover, we take the chemical potential $\mu$ to lie in this gap, $E_3(\vb{k})>\mu>E_{1, 2}(\vb{k})$, so that the bands $\ket{u_1}$ and $\ket{u_2}$ are occupied.

The set of equivalence classes of topologically similar Hamiltonians of this type may be established by examining the classifying space, given by the real Grassmannian
\begin{equation}
    \mathsf{Gr}_{2, 3}(\mathbb{R}) = \frac{\mathsf{O}(3)}{\mathsf{O}(2)\times \mathsf{O}(1)}\cong \mathbb{R}P^2.
\end{equation}
where $\mathbb{R}P^2\cong S^2/\mathbb{Z}_2$ is the real projective plane. For our purposes, it is sufficient to replace this non-orientable space with its oriented double cover $\mathsf{Gr}^+_{2, 3}(\mathbb{R})\cong S^2$. The topological phases of this system are then characterized by the distinct homotopy classes of maps between the BZ and the classifying space. The set of such maps does not form a group, and is in fact given by the set
\begin{align}\label{eq:MapClasses}
\begin{split}
    [T^3, S^2] ={}& \Big\{(v_0; \vb{v})\, \Big| \,  \vb{v}=(v_1, v_2, v_3)\in \mathbb{Z}^3; \\
    &\quad v_0\in \begin{cases} \mathbb{Z} & \vb{v}=\vb{0}\\ \mathbb{Z}_{2\gcd(\vb{v})} & \text{otherwise.}\end{cases}\Bigg\}\bigg\}
\end{split},
\end{align}
where $\gcd(\vb{v})$ is the greatest common divisor of the integers $v_1$, $v_2$ and $v_3$. We will now demonstrate that the index $v_0$ corresponds to the `strong' Hopf invariant of $\hh$, while the vector $\vb{v}$ labels its `weak' Euler invariants on each of the coordinate planes. 

To realize this correspondence, we note that by applying the band flattening procedure, whereby the occupied and unoccupied energy bands are adiabatically changed to $E_{3}(\vb{k})=+1$ and $E_{1, 2}(\vb{k})=-1$ respectively, we may bring any three-band Hamiltonian $H_3(\vb{k})$ into the form
\begin{align}\label{eq:3BandFlat}
\begin{split}
    \bar{H}_3(\kv) ={}& R_3(\vb{k})\diag\mqty(1, & -1, & -1) R_3(\vb{k})^\text{T}\\
    ={}& 2 \dh(\kv) \otimes \dh(\kv)^\text{T} - \mathbbm{1}_3,
\end{split}
\end{align}
which we refer to as the flattened Hamiltonian. Here $\dh(\vb{k})=\vb{u}_3(\vb{k})$ is the (normalized) third eigenvector with energy $+1$, and ${R_3(\vb{k})=(\ket{u_3}, \ket{u_2}, \ket{u_1})}$ is an $\mathsf{SO}(3)$ matrix with columns given by of the eigenvectors of $H_3(\vb{k})$ (the vectors have been ordered for later convenience). In particular, the vector $\dh:T^3\to S^2$ explicitly gives the map to the sphere which specifies the topological class of the Hamiltonian. Hence, the problem of determining the topological phase realized by a Hamiltonian $H_3(\vb{k})$ is reduced to finding which class in $[T^3, S^2]$ the map $\dh$ belongs to. Away from the flat band limit, it is not possible for the Hamiltonian to be expressed directly in terms of the winding vector $\dh$ as in Eq.~\eqref{eq:3BandFlat}. Nonetheless, all the following formulas involving $\dh$ may be applied directly in this case by using the third eigenvector $\vb{u}_3$ instead.

As described in Ref.~\cite{Kennedy_2016}, the topological class realized by $\dh$ is uniquely determined by its framed preimage, which may be thought of as the `ribbon' defined by the preimages of two infinitesimally separated points on the sphere. This correspondence is realized via the Pontryagin-Thom construction (see App.~\ref{app::A}), which shows that two maps $\dh_1$ and $\dh_2$ are in the same topological class if and only if their framed preimages are framed cobordant.~As we will shortly clarify, this construction provides a means by which the topological phase of the system can be deduced by inspecting the preimages of any two points on the sphere. In brief, the preimage of a point on the sphere under this map is a one-dimensional subset of the BZ composed of (contractible and/or non-contractible) loops, and the phase is determined by the types of structures (e.g. links) that are formed by two such preimages. Interestingly, these structures are not merely a computational tool: as we demonstrate in Sec.~\ref{sec::III}, in Hopf-Euler phases they can be realized physically as nodal lines.

\begin{figure}
\centering
\def\svgwidth{.47\textwidth}
\begingroup%
  \makeatletter%
  \providecommand\color[2][]{%
    \errmessage{(Inkscape) Color is used for the text in Inkscape, but the package 'color.sty' is not loaded}%
    \renewcommand\color[2][]{}%
  }%
  \providecommand\transparent[1]{%
    \errmessage{(Inkscape) Transparency is used (non-zero) for the text in Inkscape, but the package 'transparent.sty' is not loaded}%
    \renewcommand\transparent[1]{}%
  }%
  \providecommand\rotatebox[2]{#2}%
  \newcommand*\fsize{\dimexpr\f@size pt\relax}%
  \newcommand*\lineheight[1]{\fontsize{\fsize}{#1\fsize}\selectfont}%
  \ifx\svgwidth\undefined%
    \setlength{\unitlength}{279.54900066bp}%
    \ifx\svgscale\undefined%
      \relax%
    \else%
      \setlength{\unitlength}{\unitlength * \real{\svgscale}}%
    \fi%
  \else%
    \setlength{\unitlength}{\svgwidth}%
  \fi%
  \global\let\svgwidth\undefined%
  \global\let\svgscale\undefined%
  \makeatother%
  \begin{picture}(1,0.48678407)%
    \lineheight{1}%
    \setlength\tabcolsep{0pt}%
    \put(0,0){\includegraphics[width=\unitlength,page=1]{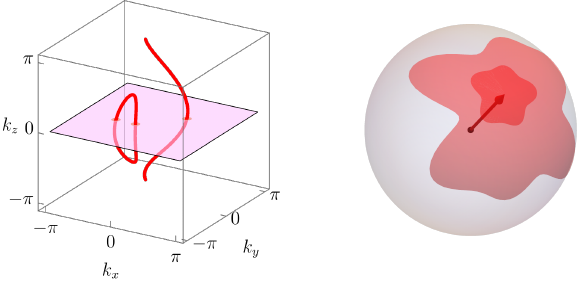}}%
    \put(0.5372629632,0.28752158){\color[rgb]{0.15686275,0.04313725,0.04313725}\makebox(0,0)[lt]{\lineheight{1.25}\smash{\begin{tabular}[t]{l}$\hat{\vb{d}}$\end{tabular}}}}%
    \put(0.52173309,0.24188894){\color[rgb]{0.15686275,0.04313725,0.04313725}\makebox(0,0)[lt]{\lineheight{1.25}\smash{\begin{tabular}[t]{l}$\longrightarrow$\end{tabular}}}}%
  \end{picture}%
\endgroup%

  \caption{The topological Hopf and Euler invariants of an (oriented) three-band Hamiltonian may be determined by examining the preimage of its third eigenvector as a map $\vb{u}_3=\dh: T^3\to S^2$. The (strong) Hopf invariant characterizes the behavior of loops and links within the 3D BZ, while the (weak) Euler classes are concerned with the intersection of these lines with 2D planes embedded within this space; see main text and Fig.~\ref{FigHopfEuler} for further discussion.}
\label{FigIntro}
\end{figure}
To proceed further it is helpful to express the rotation matrix $R_3$ and the winding vector $\dh$ in terms of quaternions (see App. \ref{app:Quaternion} for a review)~\cite{Unal_2020}. This is made possible by the well-known isomorphisms $\mathsf{SO}(3)\cong\mathsf{SU}(2) / \mathbb{Z}_2$ and $\mathsf{SU}(2)\cong S^3$, along with the embedding of the unit three sphere $S^3$ into the quaternion algebra $\mathbb{H}$ as the set of unit quaternions (versors), that is, 
\begin{align}
\begin{split}
    S^3\cong \mathbb{H}_0 =\{ & q=x_0+\ii x_1+\jj x_2+\kk x_3\in\mathbb{H}\, | \\ 
        &|q|^2 = \bar{q} q=x_0^2+x_1^2+x_2^2+x_3^2=1\}. 
\end{split}
\end{align}
By viewing the set of purely imaginary quaternions $\mathbb{H}^*=\{w\in\mathbb{H}\,|\, w=\bar{w}\}=\{w_1 \ii+w_2\jj+w_3\kk\,|\, w_i\in\mathbb{R}\}$ as $\mathbb{H}^*\cong\mathbb{R}^3$, it is possible to identify the action of rotation matrices on vectors in 3D with that of unit quaternions on imaginary quaternions via conjugation. In this way, the $\mathsf{SO}(3)$ matrix $R_3(\vb{k})$ may be written in terms of the action of a unit quaternion $q$ on the imaginary units ${\ii, \jj, \kk\in \mathbb{H}}$:
\begin{align}\label{eq:SO3Rot}
\begin{split}
    R_3(\vb{k}) ={}& \mqty(\ket{\bar{q}\ii q} && \ket{\bar{q}\jj q} && \ket{\bar{q}\kk q})\\
    ={}& \mqty(\ket{u_3(\vb{k})} && \ket{u_2(\vb{k})} && \ket{u_1(\vb{k})}),
\end{split}
\end{align}
where $\bar{q}w q = R_3 \vb{w}$, with $\vb{w}=(w_1, w_2, w_3)^\text{T}$. Since the winding vector is equal to the third eigenvector $\ket{u_3(\vb{k})}$, it follows that 
\begin{align}\label{eq:WindingVec}
    \begin{split}
        \bar{q}\ii q ={}& [x_0^2+x_1^2-x_2^2-x_3^2]\ii+2[x_1 x_2 - x_0 x_3]\jj\\
        & + 2[x_0 x_2+ x_1 x_3]\kk \\
        ={}& \dh\cdot\mqty(\ii & \jj & \kk),
    \end{split}
\end{align}
from which we can read off the components of the winding vector $\dh$. This may be conveniently summarized in terms of the Pauli matrices $\sigma_i$ as  $\hat{d}_i= \vb{z}^\dagger\sigma_i\vb{z}$, where $\vb{z} = (x_0+\ii x_1, x_2+\ii x_3)^\text{T}$. We give the explicit formula for $R_3$ in terms of the components $x_\mu$ in App. \ref{app:Quaternion}. Using this expression, along with Eq. \eqref{eq:WindingVec}, one may verify directly that the decomposition Eq. \eqref{eq:3BandFlat} holds for any quaternion $q$ of unit magnitude.

\begin{figure}
\centering 
\def\svgwidth{.47\textwidth}
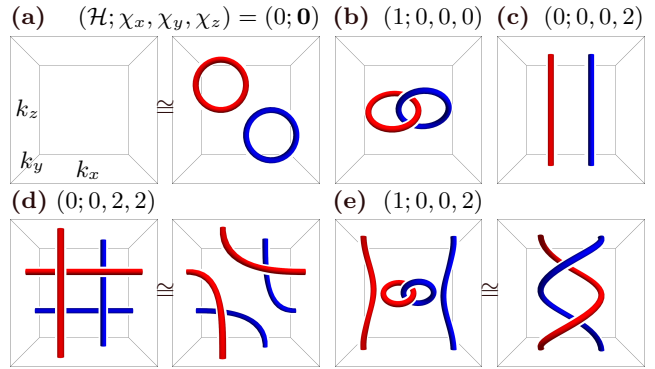
\caption{Distinct realizations of the Hopf-Euler insulator invariants, with preimages of two points on the sphere $S^2$ under the map defined by the third eigenvector $\vb{u}_3$ of $H_3$ in different topological phases. The Hopf invariant $\hh$ is equal to the linking number of the red and blue loops, while the Euler class $\chi_i$ is equal to (twice) the number of noncontractible loops  in the $i^{\text{th}}$ direction. \textbf{(a)} Trivial phase \textbf{(b)} Strong Hopf phase \textbf{(c, d)} Layered Euler phases \textbf{(e)} Hopf-Euler phase.}
\label{FigHopfEuler}
\end{figure}

We will now utilize the formalism laid out above to enumerate the possible topological phases of these Hamiltonians, which are summarized in Fig.~\ref{FigHopfEuler}.

\sect{Trivial bulk topology} First, before discussing the non-trivial topological phases that may exist in the bulk of the system, we mention the description of the trivial phase, in which all invariants vanish, in terms of the preimage construction as illustrated in Fig. \ref{FigHopfEuler}\textbf{(a)}. The trivial phase may be defined as the set of Hamiltonians for which the third eigenvector $\vb{u}_3$ is homotopic to the constant map $\dh_0(\vb{k}) = \hat{\vb{e}}_z$, the unit vector in the $z$-direction. It is straightforward to see that the preimage of a point $p$ on the sphere in this phase may be empty. However, it is also possible for $\vb{u}_3^{-1}(p)$ to consist of a finite number of unlinked, disjoint, contractible loops within the BZ, since these may be removed by a continuous transformation. Moreover, if the preimages of two points on the sphere both contain loops, then these loops also cannot link, for this would imply that $\vb{u}_3$ were not nullhomotopic. This similarly applies to each of the non-trivial phases: It is always possible to add any number of trivial loops to the preimage of a point. 

\sect{Strong Hopf topology} In the absence of weak invariants, the bulk topology of the system is classified by the homotopy group $\pi_3[S^2] \cong \mathbb{Z}$. The Hopf invariant of a map $\dh$ can be non-zero only when it is surjective, so that every point on the sphere $S^2$ corresponds to a circular preimage within the BZ (Fig.~\ref{FigIntro}); the linking number of two such circles selected from distinct points on $S^2$ is precisely equal to the Hopf invariant, see Fig. \ref{FigHopfEuler}\textbf{(b)}. This statement is true modulo homotopy -- for instance, a pair of doubly-linked loops may be deformed into two sets of singly-linked loops, which together have linking number~2. This interpretation plays an integral role in this work and we will make repeated reference to it throughout the text.

The Hopf invariant $\hh$ of a three-band model may be calculated by using the non-Abelian Berry connection of the occupied subspace, $A_{ab} = \ii\braket{u_a}{\dd u_b} = A_{ab}^i\,\dd k^i$. To do so, one must compute the Euler connection $\aa = \Pf(-\ii A) = \braket{u_1}{\partial_i u_2}\,\dd k^i$, which is equal to the off-diagonal element of the matrix $A$. The Hopf invariant is then given by,
\begin{equation}\label{eq:HopfIntegral}
    \hh = -\frac{1}{16\pi^2} \int_{T^3} \mathrm{a}~\wedge~\Eu = -\frac{1}{16\pi^2} \intBZ \dd^3\kv\, \vb{a}\cdot(\curl\vb{a}),
\end{equation}
where we have written $\aa = a_i\,\dd k^i= \vb{a}\cdot\dd\vb{k}$ and identified the Euler form
\begin{equation}\label{eq:EulerForm}
    \Eu = \dd\aa =\frac{\braket{\partial_i u_1}{\partial_j u_2} - \braket{\partial_j u_1}{\partial_i u_2}}{2}\,\dd k^i\wedge\dd k^j.
\end{equation}
To prove Eq. \eqref{eq:HopfIntegral}, we begin with the Whitehead formula for $\hh$, which reads \cite{Nakahara}
\begin{equation}\label{eq:Whitehead}
    \hh = -\frac{1}{4\pi^2}\int \omega\wedge\dd\omega,
\end{equation}
where $\omega =\Re[-\ii q\dd\bar{q}]$. As we shown in App.~\ref{app::B}, the Euler connection $\aa$ may be expressed in terms of $\omega$ as $\aa = 2\omega$, which then gives the required result. Notably, the factor of $2$ arises from the quadratic dependence of $\vb{d}(\kv)$ on $z$; this is in contrast to the four-band case (discussed in Ref. \cite{Lim2023} and Sec. \ref{sec:FourBandPhases}), where each of the vectors $\nh_\pm$ depend only linearly on each of the quaternions $q_+$ and $q_-$ (though the Hamiltonian $H_4$ is still quadratic in $q_\pm$).

The Hopf invariant may also be calculated directly from the normalized complex vector $\hat{\textbf{z}}$ provided by the decomposition $\hat{d}_i=\vb{z}^\dagger\sigma_i\vb{z}$ of the winding vector $\dh$ as
\begin{equation}\label{eq:HopfZ}
    \hh = -\frac{1}{4\pi^2}\intBZ \dd^3\vb{k}~ \varepsilon_{ijk} \hat{\textbf{z}}^{\dagger}(\partial_{i} \hat{\textbf{z}}) (\partial_{j} \hat{\textbf{z}}^{\dagger})(\partial_{k} \hat{\textbf{z}}).
\end{equation}
If the vector $\vb{z}$ is known then Eq. \eqref{eq:HopfZ} provides the quickest route for computing $\hh$~\cite{Unal2019}, as it does not require the calculation of any intermediate quantities (such as $\Eu$). However, if one knows only $\dh$, then determining $\vb{z}$ requires the solution of a partial differential equation, and it is significantly more straightforward to employ Eq.~\eqref{eq:HopfIntegral} instead \cite{Hopf_1, Lim2023}.

\sect{Layered Euler topology} In addition to the bulk index $\hh$, the Hamiltonian $H_3(\vb{k})$ may also possess codimension-1 topological quantum numbers on each of the three coordinate planes, see Fig. \ref{FigHopfEuler}\textbf{(c, d)}. We consider first the scenario in which the Hopf invariant vanishes but one or more of these weak invariants are non-zero. For concreteness, in the following we consider the plane $Q(k_z)\cong T^2$ defined by fixing $k_z$, but identical arguments apply to the $k_x$- and $k_y$-planes also. The Euler form Eq. \eqref{eq:EulerForm} may then be used to compute the Euler class $\chi_z$ of $Q(k_z)$,
\beq{eq:EulerInt}
    \chi_z = \frac{1}{2\pi} \int_{T^2_z} \mathrm{Eu} = \frac{1}{2\pi} \int_{Q(k_z)} \dd^2 \kappa\,\grad_{\KK}\cross\vb{a},
\eeq
where $\KK = (k_x, k_y)$. In particular, the Euler curvature form $\mathrm{Eu}$, as well as the Euler class itself, can be efficiently computed numerically using a band complexification trick, as introduced in Ref.~\cite{Bouhon_2019}.
This invariant has been extensively investigated in two dimensions~\cite{bouhon2018wilson, PhysRevX.9.021013, Bouhon_2019, Bouhon2020_geo, Unal_2020}, where its relation to the band degeneracies present in the relevant band subspaces was explored in detail~\cite{PhysRevX.9.021013, Bouhon2020_geo, Jiang_2021}. Additionally, when the invariant-hosting two-band subspace is not isolated with band gaps, as in semimetals, the Euler class invariant can be extended to two-dimensional patches $\mathcal{D}$ within BZ ($\mathcal{D} \in \text{BZ}$) that exclude band degeneracies residing between the other bands. Specifically, the patch Euler class is defined by including a boundary term as~\cite{Bouhon_2019, Jiang_2021}, 
\beq{}
    \chi_\mathcal{D} = \frac{1}{2\pi} \int_{\mathcal{D}} \mathrm{Eu} - \frac{1}{2\pi} \int_{\mathcal{\partial D}} \vb{a}.
\eeq
Physically, the patch Euler class $\chi_\mathcal{D}$ quantifies the stability of nodes to annihilation~\cite{PhysRevX.9.021013, Bouhon_2019, Bouhon2020_geo, Jiang_2021}. Finally, we note that the Euler invariant itself can be probed using signatures in the quench dynamics~\cite{Unal_2020}, as  was recently experimentally demonstrated in trapped-ion quantum simulators~\cite{Zhao_2022}.

The Euler class is equal to (twice) the topological degree of the unit vector $\hat{\vb{w}}_{z}(\boldsymbol{\kappa}) = \dh(\vb{k})|_{k_z=\text{const.}}$ when considered as a map $\hat{\vb{w}}:T^2\to S^2$~\cite{Unal_2020}. Indeed, $\grad_{\KK}\cross\vb{a}=\hat{\vb{w}}\cdot(\partial_x\hat{\vb{w}}\cross\partial_{y}\hat{\vb{w}})$ is equal to the skyrmion density of $\hat{\vb{w}}$. $\chi_z$ may alternatively be computed by counting the signed number of points in the preimage of any regular point $p\in S^2$:
\begin{equation}\label{eq:EulerDet}
    \chi_z = 2\sum_{\boldsymbol{\kappa}_p\in \hat{\vb{w}}^{-1}(p)} \sgn \det D \hat{\vb{w}}|_{\KK = \KK_p},
\end{equation}
where $D\hat{\vb{w}}$ is the Jacobian matrix. In particular, a non-zero Euler class implies the existence of a non-empty preimage $\vb{w}^{-1}(p)\subset Q(k_z)$ for all $p\in S^2$. The Euler class on the plane $Q(k_z)$ can change as a function of $k_z$ only if there is a gap closing between the conduction and valence bands at some value of $k_z$. Since we are solely concerned with gapped phases here, we exclude this possibility, hence the Euler class must remain constant for all $k_z$. It therefore follows that the preimage $\hat{\vb{w}}_{z}^{-1}(p)$ exists and is continuous for all $k_z$, so that $\dh^{-1}(p)$ is a set of lines connecting the $k_z=\pm\pi$ surfaces. This relation is shown schematically in Fig. \ref{FigHopfEuler}\textbf{(c)}. 

In general, the layered Euler phases are characterized by a triple $\boldsymbol{\chi} = (\chi_x, \chi_y, \chi_z)\in(2\mathbb{N})^3$. A representative preimage for a phase with Euler class $\boldsymbol{\chi}$ has $\chi_i/2$ non-contractible loops along the $i^{\text{th}}$ direction of the BZ for $i=x, y, z$ [see Fig. \ref{FigHopfEuler}\textbf{(d)}].

\sect{Hopf-Euler topology} As outlined above, the Hopf invariant $\hh$ of the map $\dh:T^3\to S^2$ is non-trivial only when the preimage of any two points on $S^2$ forms a link in the BZ. Similarly, the Euler class $\chi$ of a coordinate plane $Q\cong T^2$ within the BZ is non-zero when the preimage of a point on $S^2$ forms a loop around the non-contractible direction of the torus perpendicular to Q, see Fig. \ref{FigHopfEuler}\textbf{(e)}. It therefore follows that if a 3-band Hamiltonian $H_3(\vb{k})$ carries both an Euler class on $Q(k_z)$ and a Hopf invariant, the preimage under $\dh$ of two points in $S^2$ must both (\textit{i}) form a link, and (\textit{ii}) connect the $k_z=\pm\pi$ planes. As we show in Fig.~\ref{FigHopfEuler}, this may be realized either as a disjoint connection of a link and two lines, or equivalently as a helix within a single BZ. 

The topological invariants of this phase may again be calculated using Eqs. \eqref{eq:HopfIntegral}, \eqref{eq:HopfZ}, \eqref{eq:EulerInt}, and \eqref{eq:EulerDet}. However, it is important to recall Eq. \eqref{eq:MapClasses}, which indicates that the presence of non-trivial weak Euler invariants leads to a reduction in the range of values which the Hopf invariant $\hh$ can take. In this way, the Hopf invariant given by Eqs. \eqref{eq:HopfIntegral} and \eqref{eq:HopfZ} must be interpreted $\mathrm{mod} (\gcd\boldsymbol{\chi})$, as can be shown with the Pontryagin-Thom construction~\cite{Kennedy_2016}; see App.~\ref{app::A} (note that the factor of 2 premultiplying the $\gcd$ in Eq.~\eqref{eq:MapClasses} is absorbed into the conventional factor of 2 in the Euler class). This should also be taken into account when inspecting the preimage of two points on $S^2$: a pair of loops with linking number $\gcd\boldsymbol{\chi}$ may be trivialized without closing the gap.

\subsection{Three-band flag phases}\label{sec:3bandflag}
So far, we have only considered phases with a single gap between the occupied and unoccupied states. It is instructive to consider the modifications to the conclusions of the previous section which occur when the more stringent condition that all phases are fully gapped is imposed. That is, we now require that $E_3(\vb{k})>\mu > E_2(\vb{k})>E_1(\vb{k})$ for all $\vb{k}\in \BZ$. The classifying space of the system in this case is given by the flag manifold, 
\begin{equation}
    \mathsf{Fl}_{1, 1, 1}(\mathbb{R}) = \frac{\mathsf{O}(3)}{\mathsf{O}(1)\times\mathsf{O}(1)\times\mathsf{O}(1)}, 
\end{equation}
where $\mathsf{O}(1)\cong\mathbb{Z}_2$. This space has homotopy groups $\pi_3[\mathsf{Fl}_{1,1,1}(\mathbb{R})] \cong \mathbb{Z}$ and $\pi_2[\mathsf{Fl}_{1,1,1}(\mathbb{R})] \cong 0$, which respectively label the strong and weak topological invariants of the Hamiltonian. Notably, the condition that the lower gap remain open forces all Euler classes to be zero. Seen from another perspective, a non-zero Euler class protects the nodes in the occupied two-band subspace from gapping out. This observation is of especial importance for Sec. \ref{sec::III}, where we will demonstrate that a non-trivial Euler class is required to stabilize the nodal helices that emerge in $\mathcal{C}_{2z}$-symmetric Hopf-Euler phases. 

A representative 3-band Hamiltonian $H_3^{\text{flag}}$ of a flag phase may be obtained by flattening the bands to $E_3(\vb{k})\mapsto +1$, $E_2(\vb{k})\mapsto 0$, and $E_1(\vb{k})\mapsto -1$. This has the effect of modifying the central diagonal matrix in Eq. \eqref{eq:3BandFlat} from $\diag(1, -1, -1)$ to $\diag(1, 0, -1)$, so that
\begin{equation}\label{eq:3BandFlag}
    \bar{H}^{\text{flag}}_3(\kv) = V_3(\vb{k})\diag\mqty(1, & 0, & -1) V_3(\vb{k})^\text{T},
\end{equation}
where $V_3(\vb{k})\in\mathsf{SO}(3)$. In contrast to Eq. \eqref{eq:3BandFlat}, it is not possible to write the flag Hamiltonian in Eq. \eqref{eq:3BandFlag} in terms of a single three-dimensional winding vector. Thus, the topology of this system may be described only with reference to the matrix $V_3$. In other words, the topological index of this system is \textit{not} an element of the group $\pi_3[S^2]$, but rather of 
\begin{align}\label{eq:FlagHomotopy}
    \pi_3[\mathsf{Fl}_{1, 1, 1}] \cong \pi_3[\mathsf{SO}(3)]\cong \pi_3[\mathsf{SU}(2)]\cong\pi_3[S^3]\cong\mathbb{Z},
\end{align}
where we have noted that higher homotopy groups are insensitive to the presence of discrete quotients, and made use of the isomorphisms $\mathsf{SO}(3)\cong\mathsf{SU}(2)/\mathbb{Z}_2$ and $\mathsf{SU}(2)\cong S^3$. An explicit expression for the topological index realized by the map $V_3:T^3\sim S^3\to \mathsf{SO}(3)$ may be found by making use of the homomorphism $f:\mathsf{SO}(3)\to \mathsf{SU}(2)$ as follows. Firstly, we note that the $\mathsf{SO}(3)$ matrix $V_3$ may be written in terms of the generators of $\mathfrak{so}(3)\cong\mathfrak{su}(2)$ as $V_3(\vb{k}) = \exp(\ii \boldsymbol{\theta}(\vb{k})\cdot\vb{L})$, where $\boldsymbol{\theta}(\vb{k})$ is a real 3-component vector of parameters. The $\mathsf{SU}(2)$ matrix corresponding to $V_3$ under $f$ is then given by $U(\vb{k}) = \exp(\ii \boldsymbol{\theta}(\vb{k})\cdot\boldsymbol{\sigma}/2)$, since the matrices $\sigma_i/2$ generate $\mathfrak{su}(2)$. The winding number of this matrix as a map $U:S^3\to \mathsf{SU}(2)$ is given by the Pontryagin index \cite{Nakahara,Zee2010-ZEEQFT},
\begin{align}\label{eq:SU2Theta}
    w ={}& \frac{1}{24\pi^2} \intBZ \text{Tr} [(U^{-1}\dd U)^3]\\
    ={}& \frac{1}{24\pi^2} \intBZ\dd^3\vb{k}\, \epsilon^{ijk}\text{Tr}~[(U^{-1}\partial_i U)(U^{-1}\partial_j U)(U^{-1}\partial_k U)]\nonumber\\
    ={}& \frac{1}{16\pi^2}\intBZ\dd^3\vb{k}\, \mathrm{sinc}^2\left(\frac{\theta}{2}\right) \partial_x\boldsymbol{\theta}\cdot(\partial_y\boldsymbol{\theta}\cross \partial_z\boldsymbol{\theta})\nonumber,
\end{align}
where $\theta(\vb{k})=|\boldsymbol{\theta}(\vb{k})|$; from Eq. \eqref{eq:FlagHomotopy} this is also equal to the homotopy class of the flag Hamiltonian $H^{\text{flag}}_3$. 

While Eq. \eqref{eq:SU2Theta} in principle allows the computation of the topological index of $H^{\text{flag}}_3$, in practice it is difficult to determine the parameters $\boldsymbol{\theta}(\vb{k})$ from the Hamiltonian. In fact, there is a more simple expression for $w$ which moreover relates it to the topological invariants of the previous section. To arrive at this formula, we note that by computing the matrix exponential $V_3=\exp(\ii \boldsymbol{\theta}\cdot\vb{L})$ we can obtain explicit expressions for the eigenvectors $\ket{u_a(\vb{k})}$ of the Hamiltonian in terms of $\boldsymbol{\theta}$. Thus, the Euler connection $\aa = \braket{u_1}{\dd u_2}$ and Euler form $\mathrm{Eu} = \dd\aa$ may be explicitly computed in terms of the parameters $\boldsymbol{\theta}$. In terms of these objects, one finds that
\begin{equation}\label{eq:SU2Winding}
    w = -\frac{1}{16\pi^2}\intBZ \aa\wedge\Eu.
\end{equation}
This allows the Pontryagin index to be computed from the eigenvectors of $H^{\text{flag}}_3$ directly, without it being necessary to find $\boldsymbol{\theta}(\vb{k})$ as an intermediate step. Equation \eqref{eq:SU2Winding} may be verified by evaluating the right hand side of the equation as a function of $\boldsymbol{\theta}$, and checking that it agrees with Eq. \eqref{eq:SU2Theta}.
We stress that, while the right hand side of Eq. \eqref{eq:SU2Winding} is identical to that of Eq. \eqref{eq:HopfIntegral}, the Pontryagin index is \textit{not} a Hopf index: the particular type of topological number that the expression $-\frac{1}{16\pi^2}\int\aa\wedge\Eu$ computes is different in the cases in which the occupied bands touch, and when they are fully gapped. Nevertheless, the fact that both of these expressions agree precisely is no coincidence, and this equality demonstrates that opening a gap in the occupied subspace causes the Hopf index (which characterizes the winding of $S^3$ around $S^2$) to become a Pontryagin index (which characterizes the winding of $S^3$ around $\mathsf{SU}(2)\cong S^3$). Another perspective on the similarity of these expressions may be obtained by noting that $\aa\wedge\Eu$ is, up to multiplication by a constant, the only gauge-invariant 3-form that exists in this system \cite{Lim2023}. Hence, given that a $\mathbb{Z}$-valued three-dimensional topological index exists both when the occupied bands touch, and when they are gapped, it is necessary that both must be proportional to the integral of $\aa\wedge\Eu$ over the BZ. 

Finally, we note that there is no way to determine the topological class of a flag Hamiltonian using a preimage construction like that described in the previous section. The Pontryagin index is equal to the degree of the smooth map $S^3\to S^3$ defined by the Hamiltonian and, since the target and base spaces of such maps have the same dimension, the preimage of a point on $S^3$ is not a loop in the BZ, but rather a discrete collection of points. Hence, there is no way to interpret the Pontryagin index as a linking number.

\subsection{Four-band Hopf-Euler phases}\label{sec:FourBandPhases}
The classification of four-band Hopf-Euler phases in three dimensions may be carried out in much the same way as was done for three-band phases. This follows by virtue of the fact that, while $\mathsf{Gr}^+_{2, 3}(\mathbb{R})\cong S^2$, the (oriented cover of the) classifying space of a half-filled four-band Hamiltonian $H_4(\vb{k})$ is given by
\begin{equation}
    \mathsf{Gr}^+_{2, 4}(\mathbb{R}) \cong \frac{\mathsf{SO}(4)}{\mathsf{SO}(2)\times\mathsf{SO}(2)}\cong S^2_-\times S^2_+,
\end{equation}
so that the topology of $H_4$ is classified by maps into a \textit{pair} of spheres \cite{Bouhon2020_geo}. This result was utilized in Ref.~\cite{Lim2023} to analyze the strong topology of $H_4$, which corresponds to the Hopf invariants $\hh_\pm\in\pi_3[S^2_\pm]$ of each sphere. In fact, the full set of topological indices classifying $H_4$ consists not only of these two Hopf $\hh_{\pm}$ invariants, but also an additional six Euler invariants $\chi_{i\pm}\in 2\mathbb{N}$ ($i=x, y, z$). As discussed in Sec. \ref{sec:3BandPhases}, the range of allowed values of the Hopf invariants is reduced in the presence of weak Euler classes, taking values $\hh_\pm\in\mathbb{Z}_{\gcd(\boldsymbol{\chi}_\pm)}$ in general. 

To compute the topological indices of a four band system, one can flatten the bands $E_{1, 2}(\vb{k})\mapsto -1$ and $E_{3, 4}(\vb{k})\mapsto +1$,  after which the Hamiltonian may be written as (c.f. Eq. \eqref{eq:3BandFlat})
\begin{align}\label{eq:4BandFlat}
\begin{split}
    \bar{H}_4(\vb{k}) ={}& R_4(\vb{k})\diag\mqty(1, & 1, & -1, & -1) R_4(\vb{k})^\text{T}\\
    ={}& \nh_-(\kv)\cdot \underline{\Gamma} \cdot \nh_+(\kv),
\end{split}
\end{align}
where $\nh_\pm$ are normalized eigenvectors in $\mathbb{R}^3$,  $\underline{\Gamma}$ is an array of Dirac matrices (given in App.~\ref{app::B}), and $R_4(\vb{k}) = (\ket{u_4}, \ket{u_3}, \ket{u_2}, \ket{u_1})$ is an $\mathsf{SO}(4)$ matrix with columnns given by the eigenvectors of $H_4(\vb{k})$. Just as an arbitrary $\mathsf{SO}(3)$ matrix can be parametrized by a single quaternion of unit norm, it is always possible to write the $\mathsf{SO}(4)$ matrix $R_4(\vb{k})$ in terms of two unit quaternions $q_\pm$ as
\begin{align}
\begin{split}
    R_4 ={}& \big(\ket{\overline{q}_- q_+}, \,\ket{\overline{q}_-\ii q_+}, \,\ket{\overline{q}_- \jj q_+},\, \ket{\overline{q}_- \kk q_+}\big)\\
    ={}& \big(\ket{u_4(\vb{k})},\, \ket{u_3(\vb{k})} ,\, \ket{u_2(\vb{k})} ,\, \ket{u_1(\vb{k})}\big).
\end{split}
\end{align}
The winding vectors may then be expressed in terms of these quaternions as
\begin{equation}
    \nh_{\pm} = \overline{q}_\pm \ii q_\pm.
\end{equation}
From here, the computation of the topological invariants is simply a matter of applying Eqs. \eqref{eq:Whitehead} and \eqref{eq:EulerInt} to each of the maps into the spheres $S^2_\pm$. Firstly, the Euler class on the $k_z=k_0$ (with $k_0$ $\text{const.}$) coordinate plane may be written as
\begin{align}\label{eq:Euler4band}
\begin{split}
    \chi_{z\pm} ={}& \frac{1}{4\pi}\int_{Q(k_z)} \dd^2\boldsymbol{\kappa}\,\nh_{\pm}\cdot(\partial_x\nh_{\pm}\cross\partial_y\nh_\pm)\\
    ={}& \frac{1}{4\pi}\int_{Q(k_z)} (\Eu^{\text{c}}\mp \Eu^{\text{v}}),
\end{split}
\end{align}
where $\boldsymbol{\kappa}=(k_x, k_y)$ and $\Eu^{\text{c,v}}=\dd\aa^{\text{c,v}}$ are the Euler forms of the valence (v) and conduction (c) subspaces respectively (analogous expressions hold for the $k_x$- and $k_y$-planes). It should be noted that, in contrast to the three-band case, these quantities do \textit{not} correspond directly to the topological quantum numbers of the occupied and unoccupied subspaces. Instead, by taking the sum and difference of the Euler classes in Eq. \eqref{eq:Euler4band}, one finds~\cite{Lim2023},
\begin{subequations}
\begin{align}
    &\chi_{i}^{\text{v}} = \frac{1}{2\pi}\int_{Q(k_i)} \Eu^{\text{v}} = \chi_{i+} + \chi_{i-},\\
    &\chi_{i}^{\text{c}} = \frac{1}{2\pi}\int_{Q(k_i)} \Eu^{\text{c}} = \chi_{i+} -\chi_{i-}.
\end{align} 
\end{subequations}
On the other hand, as a result of the delicate, or unstable, bulk topology it is not possible to separate the Hopf invariants $\hh_\pm$ in this way \cite{PhysRevB.102.165151,alexandradinata2021}. The Hopf invariants can be computed from the Whitehead formula for the maps on each sphere:
\begin{align}\label{eq:HopfPM}
\begin{split}
    \hh_{\pm} ={}& -\frac{1}{4\pi^2}\int \omega_{\pm}\wedge\dd\omega_{\pm}\\
    ={}& -\frac{1}{16\pi^2}\intBZ (\aa^{\text{c}}\mp\aa^{\text{v}})\wedge(\Eu^{\text{c}}\mp\Eu^{\text{v}});
\end{split}
\end{align}
where $\omega_\pm = \Re[-\ii q_\pm\dd\overline{q}_\pm]$, and in the second line we used $\omega_{\pm}= (\aa^{\text{c}}\mp\aa^{\text{v}})/2$ \cite{Lim2023}. There is no way to use Eqs. \eqref{eq:HopfPM} to obtain an expression that depends exclusively on quantities from either of the occupied or unoccupied subspaces~\cite{Lim2023}. This difference can ultimately be traced to the fact that the weak invariants may be computed directly from the winding vectors $\nh_\pm$, while the strong invariants are naturally expressed in terms of the quaternions $q_\pm$ \cite{Lim2023}.

The topological phases of a real four-band model also admit a description in terms of the preimage construction. Since the classification of these phases is in terms of two winding vectors $\nh_\pm$, there are now two preimages to consider. The topological invariants $(\hh_\pm; \boldsymbol{\chi}_\pm)$ corresponding to each vector may be determined by looking at the preimage of $\vb{n}_\pm$, in exactly the same way as the invariants $(\hh; \boldsymbol{\chi})$ of a three-band phase are found from the preimage of the winding vector $\dh$ (see Sec. \ref{sec:3BandPhases}). 

Finally, we remark that, like the three-band Hopf phases, which have a fully-gapped flag phase limit with a $\mathbb{Z}$ bulk index, the four-band Hopf phases admit transitions to the fully-gapped four-band flag phases, which were introduced in Ref.~\cite{PhysRevB.109.165125}. For completeness, we elaborate on these phases in App.~\ref{app::C}.

\section{Physical manifestations}\label{sec::III}

In this section we discuss the physical manifestations of non-Abelian three-band Hopf topologies, which include \textit{(i)} a bulk quantized non-linear optical effect, \textit{(ii)} quantum-geometric breathing in the hybrid Wannier functions, \textit{(iii)} boundary states with a surface Euler invariant protected under $\mathcal{C}_2\mathcal{T}$ symmetry, and \textit{(iv)} helical nodal structures naturally realized in the presence of the non-trivial weak Euler invariants.

\subsection{Quantized shift effect}
In the following we demonstrate the existence of a quantized shift response~\cite{jankowski2024quantized} in the three-band real Hopf insulator. Here, we consider two lower bands $\ket{u_1(\vb{k})}, \ket{u_2(\vb{k})}$ of the three-band Hopf-Euler, or flag phases, both fully occupied with electrons that couple to light. The quantity of interest is the shift current $j^i_{\text{shift}}(0)$ induced on the photoexcitation of electrons, which is the second-order DC bulk photovoltaic response present due to the incidence of an AC electromagnetic field with frequency $\omega$~\cite{PhysRevB.48.11705, PhysRevB.61.5337}:
\beq{}
    j^i_{\text{shift}}(0)= 2 \sigma^{ijk}_{\text{shift}}(\om) \mathcal{E}_j(\om) \mathcal{E}_k(-\om), 
\eeq
where $\mathcal{E}_i(\om)$ are the frequency components of the electric field, and we have left the sum over the spatial indices $j, k=x, y, z$ implicit. We stress that here we consider the excitation part of the shift photocurrent, assuming that the electrons have no time for the relaxation, as can be targeted in the transient shift responses on the subpicosecond timescales~\cite{1982JETP...56..359B, zhu2024anomalousshiftopticalvorticity}. The shift photoconductivity can be written as~\cite{PhysRevX.10.041041,Ahn_2021},
\begin{small}
\beq{eq:shift}
    \sigma^{ijk}_{\text{shift}}(\om) = \frac{\pi e^3}{2} \sum_{mn} \int_{\text{BZ}} \frac{\dd^3\kv}{(2\pi)^3} \delta(\om - \om_{mn}) f_{mn} \ii\left(C^{mn}_{kij}-(C^{mn}_{jik})^* \right),
\eeq
\end{small}
where $\om_{mn} = E_{m\kv}-E_{n\kv}$ and $f_{mn} = f_{m\kv} - f_{n\kv}$ are respectively the difference in energies and Fermi occupation factors between the bands $m$ and $n$; note that we set $\hbar=1$. In the zero-temperature limit, the factors $f_{mn}$ are non-zero only when $m$ is unoccupied and $n$ is occupied, or vice versa, and in this case they are equal to $\pm$1. The coefficients $C^{mn}_{kij}$ are the components of a Hermitian connection~\cite{Ahn_2021}, and they are given by
\beq{}
    C^{mn}_{kij} = A^{k}_{mn} \nabla_i A^{j}_{nm},
\eeq
where the diagonal elements $A^{i}_{nn}$ vanish under $\mathcal{PT}$ symmetry, and the covariant derivative of the off-diagonal elements of the Berry connection is defined as
\begin{align}\label{cov}
\begin{split}
    \nabla_{i} A^{j}_{nm} ={}& \partial_{i} A^{j}_{nm} -\ii(A^{i}_{nn} - A^{i}_{mm}) A^{j}_{nm}.
\end{split}
\end{align}
Notably, in the context of photovoltaic responses, the non-Abelian Berry connection $A^{i}_{nm}$ may be naturally interpreted as a transition dipole matrix element~\cite{Ahn_2021}. Following Ref.~\cite{jankowski2024quantized}, and defining $F_{\text{sym}} \equiv {-\ii \int \dd \om~ [\sigma^{xyz}_{\text{shift}}(\om) + \sigma^{yzx}_{\text{shift}}(\om) + \sigma^{zxy}_{\text{shift}}(\om)]}$, we find that
\beq{}
    F_{\text{sym}} = \frac{2e^3}{\hbar^2} \hh.
\eeq
It should be noted that this second-order quantized integrated shift effect can be non-zero only when inversion symmetry $\mathcal{P}$ is broken. Indeed, if the $\mathcal{P}$ symmetry is preserved, then $\hh = 0$ (see App. \ref{app::D} for more details). 

Importantly, we note that the photovoltaic response of the Hopf-Euler insulators is fundamentally different from the photovoltaic response in the complex Hopf insulators, as related to a returning Thouless pump (RTP) realized by these phases~\cite{alexandradinata2021, alexandradinata2022topological} (see also App.~\ref{app::E}). The reason for such a distinction is that due to the $\mathcal{PT}$ symmetry, the linear shift photoconductivities~\cite{PhysRevX.10.041041, Ahn_2021} completely vanish in the Hopf-Euler insulators, unlike in the Hopf insulators where the linear shift response is reflected by the RTP~\cite{alexandradinata2022topological}. On the contrary, as we demonstrate here and elaborate further in App.~\ref{app::D}, the Hopf-Euler insulators support a shift response to circularly polarized light, while their shift response to linearly polarized light vanishes identically. Additionally, the associated three-band flag phase limits realize a sum rule related to the torsion tensor~\cite{jankowski2024quantized} (see also App.~\ref{app::D}), which remains vanishing in the two-band models~\cite{PhysRevX.10.041041, Ahn_2021} that describe the complex Hopf insulators supporting RTP.

\subsection{Quantum-geometric breathing}
Hopf-Euler invariants also have an effect on the behavior of the system in real space, where they are apparent in quantum-geometric bounds and breathing of the Wannier functions of the topological bands.

The presence of non-trivial weak Euler invariants on any of the coordinate planes within the BZ places bounds upon the real-space localizability of Wannier functions. To see this, we begin with a bound relating the quantum metric to the Euler class:
\beq{}
    g_{ii} + g_{jj} \geq 2 |\text{Eu}_{ij}|,
\eeq
where $i,j=x,y,z$ are spatial indices (see App. \ref{app::E}). By combining the three inequalities of this form, we find that the trace of the quantum metric is bounded from below by the weak invariants,
\beq{eq:QMBound}
    \text{Tr}~g = g_{xx} + g_{yy} + g_{zz} \geq |\mathrm{Eu}_{xy}| + |\mathrm{Eu}_{yz}| + |\mathrm{Eu}_{zx}|.
\eeq
It is well known that the trace of the quantum metric is directly related to the localization of Wannier functions in real space~\cite{PhysRevLett.82.370}. In particular, the variance of the position of a Wannier function is given by
\begin{equation}\label{eq:WannierVariance}
    \sigma_{r}^2 =\langle r^2\rangle-\langle r\rangle^2= \frac{V}{(2\pi)^3} \int \dd^3\kv~\text{Tr}~g.
\end{equation}
Combining Eqs.~\eqref{eq:QMBound} and \eqref{eq:WannierVariance} we find that,
\begin{align}
\begin{split}
    \sigma_{r}^2 \geq{}& \frac{V}{(2\pi)^3} \int \dd^3\kv~(|\mathrm{Eu}_{xy}| + |\mathrm{Eu}_{yz}| + |\mathrm{Eu}_{zx}|) \\
    \geq{}& \frac{1}{4\pi^2} (A_{x}|\chi_x|+A_{y}|\chi_y|+A_{z  }|\chi_z|), 
\end{split}
\end{align}
where $V$ is the volume of the unit cell, and $A_i = V/a_i$, with $a_i$ the lattice parameter along the $i^{\text{th}}$ coordinate direction. Here we have noted that, for example
\begin{equation}
   \int\frac{\dd^3\vb{k}}{(2\pi)^3}\,|\mathrm{Eu}_{xy}| \geq \int\frac{\dd k_z}{2\pi}\,\left|\int\frac{\dd^2 \boldsymbol{\kappa}}{(2\pi)^2}\,\mathrm{Eu}_{xy}\right| = \frac{1}{a_z} \frac{|\chi_z|}{2\pi},
\end{equation}
since the Euler class $\chi_z$ is independent of $k_z$. 

In addition to the bound argument given above, in App.~\ref{app::E} we demonstrate analytically and numerically that the maximally-localized \textit{hybrid} Wannier functions of real Hopf insulators show periodic oscillations in their second moment, that is, $\sigma^2_{r}(k_z)$ oscillates as $k_z$ is changed. In metamaterial or cold atom realizations of Hopf-Euler insulators, this could be experimentally deduced from wavefunction tomography, as discussed further in Sec.~\ref{sec::V}. 

\subsection{Boundary states}

We now address the effective theory for the boundary states of the three-band RHIs that is induced by the presence of strong Hopf invariants. We construct a continuum bulk
Hamiltonian upon introducing a domain-wall
configuration in the mass parameter ($m$) profile (see also App.~\ref{app::F}): $m(z) = A z$ for a region around $z=0$. Here $A$ determines the steepness of the domain walls, and we choose $A=1$ for simplicity. The continuum Hamiltonian $H^\text{cont}_3$ then reads
\beq{}
\begin{small}
    H^\text{cont}_3 =
    2 \textbf{d}(k_x,k_y,\partial_z) \otimes \textbf{d}(k_x,k_y,\partial_z)^\text{T} - |\textbf{d}|^2 \mathsf{1}_3,
\end{small}
\eeq
where the operator ${\textbf{d}(k_x,k_y,\partial_z) = \textbf{z}^{\dagger}_{\partial} \sigma \textbf{z}_{\partial}}$ is defined in terms of ${\textbf{z}_{\partial} \equiv (k_x + \ii k_y , -\ii \partial_z + \ii m)^\text{T}}$ for a three-band RHI vacuum.~Additionally, from the continuum bands obtained on diagonalizing $H^\text{cont}_3$, it follows that the Hopf index ${\hh = \text{sgn}(m)/2}$. Having utilized these relations (see also App.~\ref{app::F} for more details on the derivation), an effective surface Hamiltonian can be obtained on projecting the bulk Hamiltonian onto the surface states. Correspondingly, the obtained effective Hamiltonian reads~\cite{PhysRevX.9.021013,Morris_2024},
\beq{}
    H_{\text{eff}} =
    \begin{pmatrix}
         -[k_x^2+k_y^2+m(a)]^2 & -g(a)(k_x + ik_y)^{2} \\
        -g(a)(k_x - ik_y)^{2} &  -[k_x^2+k_y^2+m(a)]^2
    \end{pmatrix},
\eeq
with $m(a) = a^2-1$ and $g(a) = 4a^2e^{-2 a^2}$. With $a \rightarrow 0^+$, in the proximity of the boundary, $H_{\text{eff}}$ corresponds to two occupied surface states with the surface invariant $\chi_s = 1$, residing at the boundary of the topological insulator with $\hh = 1$, and demonstrating that, here, we obtain $\chi_s = \hh$. We note that such a relation is similar to the correspondence of the surface Chern numbers $C_s = \hhc$ at the boundaries of the complex Hopf insulator~\cite{alexandradinata2021}. Moreover, analogously to the teleportation of Berry curvature in the two-band complex Hopf insulator~\cite{alexandradinata2021}, the bulk transition to a topological phase of the three-band RHI can be viewed as a teleportation of the Euler curvature. It should be noted that $\chi_s$, and equivalently the surface Wilson loop windings, are protected by a $\mathcal{C}_2\mathcal{T}$ symmetry on the surface, enforcing a reality condition. To validate the analytical argument, we include both bulk and surface Wilson loops in App.~\ref{app::F}. Once the symmetry is broken, and the time-reversal symmetry is absent, the surface Chern bands with $C_s = \hh$ can be obtained on the surfaces, as descendants of the fragile invariant $\chi_s \rightarrow C_s$~\cite{bouhon2022multigap}. By extending a similar argument to four-band Hamiltonians (see App.~\ref{app::F}), our finding is consistent with the result for the four-band RHI phases~\cite{Lim2023}, when $\mathcal{C}_2\mathcal{T}$ symmetry is absent on the surface. Moreover, we remark that the surface Euler class could be trivialized by, e.g. attaching a two-dimensional Euler insulator to the surface, and allowing the surface Euler states to hybridize with the added Euler bands. Such a scenario is similar to the case of the integer surface Chern numbers and associated surface anomalous Hall conductivities that can be annihilated by introducing additional two-dimensional Chern insulators at the surface. Hence, we expect that the relation between the bulk Hopf invariant and the surface Euler invariant can break down, even if the $\mathcal{C}_2\mathcal{T}$ symmetry was preserved on the surface, as soon as the additional bands hybridize with the surface Euler states, which can either trivialize the surface invariant, or reduce the surface Euler class to the second Stiefel-Whitney invariant~\cite{Bouhon2020_geo}. We stress here that rather than demonstrating a general bulk-boundary correspondence, we only obtain an analytical relation between the bulk and surface theories in particular models of the three-band real Hopf insulators. We finally conclude by noting that while the exact bulk and surface theories provided in our work reflect the parametrization of specific models, we would expect the analogous relation to hold in other models, similarly to the complex Hopf insulators~\cite{alexandradinata2021}. Correspondingly, we numerically retrieve that $\chi_s = \mathcal{H}$ in a set of other three-band RHI models with $\mathcal{H} > 1$, as we directly demonstrate with an interplay of bulk and surface Wilson loops in App.~\ref{app::G}; see Figs.~\ref{figWilson},~\ref{figBBC}.

\subsection{Nodal helices}
As explicitly demonstrated in the previous section, the preimage construction provides a natural picture for the interplay of the strong Hopf and weak Euler invariants in Hopf-Euler insulators. This is most clearly seen in the helical lines in the three-dimensional BZ, shown in Fig.~\ref{FigHopfEuler}, which are present only when both types of invariant are nontrivial. In fact, it is not only the preimage that can form a helical shape in this case. We now elaborate on the observation that Hopf-Euler insulators may support nodal helices within the occupied subspace. More precisely, the locus of points defined by $E_1(\vb{k})=E_2(\vb{k})$ naturally lies in the same homotopy class as such a preimage. We explicitly demonstrate this direct correspondence in Figs.~\ref{Fig:Helix} and \ref{fig:4bandModels}. 

We stress that, while the presence of the nodal lines is certainly protected by the weak Euler invariants (this is a result of the application of the Poincar\'e-Hopf index theorem~\cite{Bouhon2020_geo} to the Euler topology), the \textit{linking} of the nodal loops in the proposed model realizations of the Hopf-Euler topologies (see Sec.~\ref{sec::IV}) could be an artefact specific to these models. This situation would be similar to the nodal structures in the $\mathcal{PT}$-symmetric three-dimensional phases with a Pontryagin index~\cite{PhysRevB.109.165125}. Quantifying the specific protection, or designing a protocol that may be used to trivialize the invariants without also removing the linking structure of the nodes, therefore poses an interesting future pursuit. In the case of the Pontryagin index, the strong topological invariant does not provide any topological protection of the nodal structure. However, it remains a possibility that topological protection of nodal helices in Hopf-Euler insulators could be guaranteed by the presence of additional symmetry constraints.

An analogous correspondence between the preimage and nodal structures is also observed in systems possessing a bulk Hopf invariant only, but no Euler class in any direction. If there are nodal lines in the occupied subspace, then these lines naturally form circular links, as shown in Fig.~\ref{Fig:Helix}\textbf{(a)} and Fig.~\ref{Fig:Helix}\textbf{(d)}. However, the presence of the Euler class is essential for preventing a gap from opening between the two occupied bands~\cite{Bouhon2020_geo}. In its absence, the nodal lines may therefore be removed entirely, for example by contracting the linked loops of the opposite charge to a point, thereby resulting in a transition to the three-band flag phase described in Sec.~\ref{sec::II}. 

We finally note that in flag phases, which are classified by a single Pontryagin index, no nodal lines are present, and therefore the correspondence between the preimage of the winding vector and the nodal lines does not hold. However, as discussed at the end of Sec. \ref{sec:3bandflag}, in a flag phase the preimage of a point in $S^3$ is in general a discrete collection of points in the BZ. Thus it appears natural that the correspondence between the preimage and the nodal lines should break down in this phase.

\section{Representative models}\label{sec::IV}
Having introduced the physical consequences, such as the quantized bulk shift effect and the nodal helices, which are provided by the Euler and Hopf invariants, we now present momentum space descriptions of tight-binding models in which these topological phases are realized explicitly.~As discussed in Sec.~\ref{sec::V}, these simple models could be experimentally simulated in cold atom systems or metamaterials.

\subsection{Flattened three-band models}\label{sec:3bandFlatModels}
We firstly provide representative flat-band Hamiltonians for each of the non-trivial topological phases discussed in Sec. \ref{sec:3BandPhases}, namely the strong Hopf, layered Euler, and Hopf-Euler phases. Since Hamiltonians of this kind are entirely specified by the winding vector $\dh(\vb{k})$ (see Eq. \eqref{eq:3BandFlat}), we give only this vector in each case. It should be noted that the $\vb{d}_\alpha$, $\alpha=1, 2, 3$, given here need not be normalized when used in Eq. \eqref{eq:3BandFlat} in order to produce a Hamiltonian in the correct topological class. However, they should be normalized when used to compute topological invariants. In each model, the winding vector depends upon a single parameter $m$, in addition to its momentum dependence. The topological phases are realized when $m=1$, and the systems can all be tuned to the trivial phase by setting $m=2$.

\begin{figure}
\centering
\begingroup%
  \makeatletter%
  \providecommand\color[2][]{%
    \errmessage{(Inkscape) Color is used for the text in Inkscape, but the package 'color.sty' is not loaded}%
    \renewcommand\color[2][]{}%
  }%
  \providecommand\transparent[1]{%
    \errmessage{(Inkscape) Transparency is used (non-zero) for the text in Inkscape, but the package 'transparent.sty' is not loaded}%
    \renewcommand\transparent[1]{}%
  }%
  \providecommand\rotatebox[2]{#2}%
  \newcommand*\fsize{\dimexpr\f@size pt\relax}%
  \newcommand*\lineheight[1]{\fontsize{\fsize}{#1\fsize}\selectfont}%
  \ifx\svgwidth\undefined%
    \setlength{\unitlength}{242.99999567bp}%
    \ifx\svgscale\undefined%
      \relax%
    \else%
      \setlength{\unitlength}{\unitlength * \real{\svgscale}}%
    \fi%
  \else%
    \setlength{\unitlength}{\svgwidth}%
  \fi%
  \global\let\svgwidth\undefined%
  \global\let\svgscale\undefined%
  \makeatother%
  \begin{picture}(1,1.61284789)%
    \lineheight{1}%
    \setlength\tabcolsep{0pt}%
    \put(0.01872003,0.49676604){\color[rgb]{0,0,0}\makebox(0,0)[lt]{\lineheight{1.25}\smash{\begin{tabular}[t]{l}\textbf{(c)}\end{tabular}}}}%
    \put(0.51740187,0.49676604){\color[rgb]{0,0,0}\makebox(0,0)[lt]{\lineheight{1.25}\smash{\begin{tabular}[t]{l}\textbf{(f)}\end{tabular}}}}%
    \put(0.01740187,0.99976569){\color[rgb]{0,0,0}\makebox(0,0)[lt]{\lineheight{1.25}\smash{\begin{tabular}[t]{l}\textbf{(b)}\end{tabular}}}}%
    \put(0.51122903,0.99976569){\color[rgb]{0,0,0}\makebox(0,0)[lt]{\lineheight{1.25}\smash{\begin{tabular}[t]{l}\textbf{(e)}\end{tabular}}}}%
    \put(0.01740187,1.55789462){\color[rgb]{0,0,0}\makebox(0,0)[lt]{\lineheight{1.25}\smash{\begin{tabular}[t]{l}\textbf{(a)}\end{tabular}}}}%
    \put(0.51122903,1.55480819){\color[rgb]{0,0,0}\makebox(0,0)[lt]{\lineheight{1.25}\smash{\begin{tabular}[t]{l}\textbf{(d)}\end{tabular}}}}%
     \put(0,0){\includegraphics[width=\unitlength,page=1]{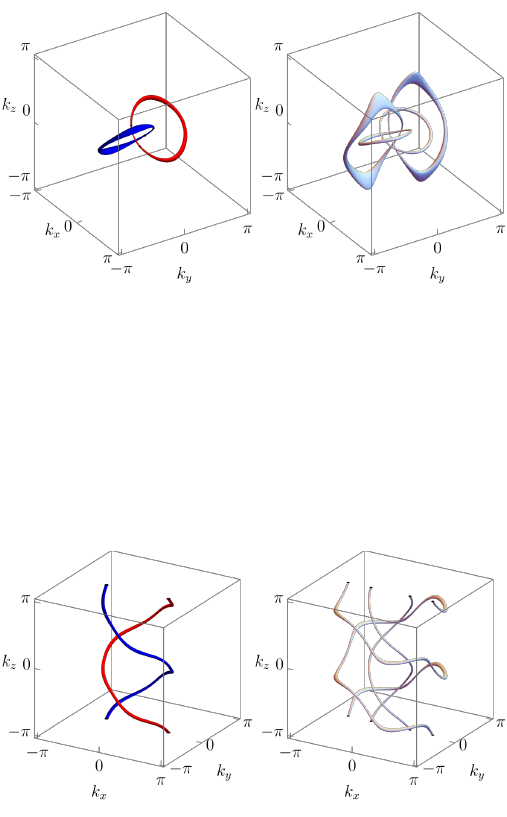}}%
     \put(0,0){\includegraphics[width=\unitlength,page=2]{PreimagesAndNodalLinesAlt.pdf}}%
  \end{picture}%
\endgroup%

  \caption{Correspondence between nodal lines and winding vector preimages in three-band Hopf/Euler insulators. \textbf{(a-c)} Preimages of the points $(1, 0, 0)$ (blue) and $(-1, 0, 0)$ (red) under the maps $\dh_\alpha:T^3\to S^2$ from Sec. \ref{sec:3bandFlatModels} with ${m=1}$. \textbf{(d-f)} Nodal lines within the occupied subspace of the corresponding perturbed Hamiltonians from Sec. \ref{sec:Dispersive3band}. \textbf{(a, d)} [Eq. \eqref{eq:3bandModelStrongHopf}] Strong Hopf phase with $\hh=1$. The linking number $L=\hh$ of the preimages gives rise to nodal links whenever the occupied bands touch, but these touchings are not protected and may be removed without closing the gap between $E_2$ and $E_3$. \textbf{(b, e)} [Eq. \eqref{eq:3bandModelLayeredEuler}] Layered Euler phase with $\vb{v}_1 = (1, 0, -2)$ and $\vb{v}_2 = (0, 1, 0)$, corresponding to $\boldsymbol{\chi} = (4, 0, 2)$. The nodal lines cannot be removed without trivializing the Euler class. {\textbf{(c,~f)}} [Eq. \eqref{Eq:3BandExample}] Hopf-Euler phase with $(\hh; \boldsymbol{\chi})=(1; 0, 0, 2)$. The nodal lines can only be removed when all weak topological indices are trivial.}
\label{Fig:Helix}
\end{figure}

\sect{Strong Hopf phase} A model with the strong non-Abelian real Hopf invariant can be realized with the winding vector
\begin{equation}\label{eq:3bandModelStrongHopf}
    \vb{d}_1(\vb{k}; m) = \textbf{z}^\dagger \sigma \textbf{z},
\end{equation}
where 
\begin{equation}
    \vb{z} = \mqty(\sin k_x + \ii \sin k_y\\
     \sin k_z + \ii \left(m+\frac{3}{2}\right)\sum_{i =x,y,z} \cos k_i ).
\end{equation}
The vector $\vb{z}$ first appeared in Ref. \cite{Hopf_1}, where it was used to construct a two-band complex Hopf insulator. 

The Hamiltonian described by this winding vector possesses a $\mathcal{C}_{2z}$ symmetry represented by the matrix $C_{2z}= \diag(-1, -1, 1)$. This is relevant for the structure of the nodal lines within the occupied subspace, which we discuss in Sec. \ref{sec:Dispersive3band}.

\sect{Layered Euler phase} To construct a representative phase hosting the weak Euler invariants $\boldsymbol{\chi}=(\chi_x, \chi_y, \chi_z)$, we mimic the procedure used in Ref. \cite{Kennedy_2016} to construct layered Chern phases. We begin with the winding vector
\begin{equation}
    \tilde{\vb{d}}_2(k_x, k_y; m) = (\sin k_x, \sin k_y, m-\cos k_x - \cos k_y)
\end{equation}
that produces a phase with Euler class $\boldsymbol{\chi}=(0, 0, 2)$. The vector $\tilde{\vb{d}}_2$ has no $k_z$ dependence, so it represents a system composed of Euler insulators stacked in the $\hat{\vb{z}}$-direction. To change the direction of this stacking to point along a given unit vector $\hat{\vb{t}}$, we choose two vectors $\vb{v}_1$ and $\vb{v}_2$ such that $\vb{v}_1\cdot\hat{\vb{t}}=\vb{v}_2\cdot\hat{\vb{t}}=0$, and we define
\begin{equation}\label{eq:3bandModelLayeredEuler}
    \vb{d}_2(\vb{k}; m) = \tilde{\vb{d}}_2(\vb{v}_1\cdot\vb{k}, \vb{v}_2\cdot\vb{k}; m).
\end{equation}
The Euler classes realized by this vector may be calculated as follows~\cite{Kennedy_2016}. Firstly, to compute $\chi_z$ we choose a plane $Q(k_z)=\{(\boldsymbol{\kappa}, k_z)\in \BZ\,|\,k_z = \text{const.}\}$, and without loss of generality take $k_z=0$. Then the winding vector $\vb{w}_z(\boldsymbol{\kappa}) = \vb{d}_2|_{k_z=0} = \tilde{\vb{w}}_z(B\boldsymbol{\kappa})$, where $B_{ij} \equiv (\vb{v}_i)_j$ with $i,j = 2$, and $\tilde{\vb{w}}_z=\tilde{\vb{d}}_2|_{k_z=0}$. Since the map $\tilde{\vb{w}}_z$ has degree 1, we can pick a regular value in $\mathsf{Gr}_{2,3}(\mathbb{R})$ with the preimage $\boldsymbol{\kappa}_0$. Then, $B \boldsymbol{\kappa} = \boldsymbol{\kappa}_0$, which yields $|\det B|$ solutions within $Q(k_{z}=0) \cong T^2$, each with orientation $\sgn \det(D \vb{w}_z) = \sgn~B$. Using Eq.~\eqref{eq:EulerDet}, it then follows that $\chi_z = 2\det B = 2(\vb{v}_1\cross\vb{v}_2)_z$, where the factor of $2$ is conventional. By repeating this argument for $\chi_x$ and $\chi_y$, we find that $\boldsymbol{\chi} = 2\vb{v}_1\cross\vb{v}_2$. Hence, like layered Chern insulators~\cite{Kennedy_2016}, any layered Euler phase can be described in an infinite number of ways through an appropriate choice of the vectors $\vb{v}_1$ and $\vb{v}_2$.

\sect{Hopf-Euler phase} Finally, we detail the winding vector for a model exhibiting a Hopf-Euler phase with $(\hh;\boldsymbol{\chi}) = (1; 0, 0, 2)$: 
\begin{equation}\label{Eq:3BandExample}
    \vb{d}_3(\vb{k}; m) = \ee^{\ii k_z L_z}\tilde{\vb{d}}_2(k_x, k_y; m),
\end{equation}
where $(L_i)_{jk} = \ii\epsilon_{ijk}$ are the generators of the $\mathfrak{so}(3)$ Lie algebra. In~Figs.~\ref{Fig:Helix}\textbf{(c)} we show the preimage of the points $\pm\hat{\vb{x}}=\pm(1, 0, 0)$ for the case $m=1$, and we verify that these lines together form a helix. The model described here may be easily extended to represent a phase with indices  $(\hh; \boldsymbol{\chi})= (p; 0, 0, 2p')$ for any integers $p, p'$, by modifying $k_z\mapsto p k_z$ and $k_x\mapsto p' k_x$.

Let us now briefly elaborate upon the procedure that we have used to construct these three-dimensional Hopf-Euler phases, namely dimensional extension~\cite{Qi_2008} from a two-dimensional Euler model. This is in direct analogy to the correspondence between Chern and Hopf-Chern insulators described in \cite{Kennedy_2016}. The matrix exponential $\ee^{\ii k_z L_z}$ in Eq.~\eqref{Eq:3BandExample} is a $k_z$-dependent rotation about the $z$-axis. Since the map $\vb{d}_2$, which describes an Euler phase layered in the $\hat{\vb{z}}$-direction, has preimages that consist of a single line connecting the $\k_z=\pm \pi$ faces of the BZ, this rotation has the effect of `twisting' these lines about the $\mathcal{C}_{2z}$-invariant line $\Gamma_z=\{\vb{k}\in\BZ\,|\, k_x=k_y=0\}$, thereby forming a helical structure in momentum space. The strands of this helix have linking number 1, with the Hopf invariant being non-trivial. Moreover, for any fixed value of $k_z$ the vector $\dh_3(k_x, k_y)$ describes a rotated Euler phase, so the Euler class in the $z$-direction is the same as the 2D model that was used to construct the 3D phase. For a summary of the dimensional extension/reduction correspondences for complex and real Hopf insulators, see Fig.~\ref{fig_dim}.

Finally, we note that, like the strong Hopf model defined above, the Hamiltonian possesses a $\mathcal{C}_{2z}$ symmetry represented by the matrix $C_{2z}= \diag(-1, -1, 1)$, which is manifestly reflected by the shape of the nodal structures, as demonstrated in~Fig.~\ref{Fig:Helix}.

\subsection{Dispersive three-band models}\label{sec:Dispersive3band}
The models given in the previous section have completely flat bands, so they do not display any of the nodal lines described in Sec. \ref{sec::III}. To move away from this degenerate limit, we add a $\mathcal{C}_{2z}$-preserving perturbation $V_3 = \lambda \diag(-1, 0, 1)$ to each Hamiltonian, giving
\begin{equation}\label{eq:3BandPerturbed}
    H_3(\vb{k}) = \bar{H}_3(\vb{k})+V_3.
\end{equation}
We have verified numerically that all of the flat-band Hamiltonians given above remain gapped for all $\vb{k}\in \text{BZ}$ provided $\lambda$ is sufficiently small; we chose the value $\lambda = 0.8$ in our computations. This perturbation lifts the degeneracy and reveals the nodal structure inherent to each of the models. In Figs. \ref{Fig:Helix}\textbf{(a-c)} we show the loci defined by $E_1(\vb{k})=E_2(\vb{k})$ for each of the perturbed models given in Sec. \ref{sec:3bandFlatModels}. In line with the discussion of Sec. \ref{sec::III}, the nodal lines in each of the perturbed models, shown in Figs. \ref{Fig:Helix}\textbf{(d-f)}, each lie within the same homotopy class as the corresponding preimages. In particular, the nodal lines in the Hopf-Euler phase form interlocking helices which, as we find numerically, are stable to $\mathcal{C}_{2z}$-preserving perturbations. The strong Hopf phase, with winding vector Eq. \eqref{eq:3bandModelStrongHopf}, has a $\mathcal{C}_{2z}$ symmetry, and we therefore expect the system to exhibit nodal links respecting the symmetry, whenever the occupied bands touch. While this is indeed the case, as shown in Fig. \ref{Fig:Helix}\textbf{(d)}, it should nonetheless be noted that these band touchings are not protected, so the nodal lines can be removed without closing the gap between $E_2$ and $E_3$.

Finally, we comment that the number of nodes is always four times the number of connected components of a single preimage. This is because a two-dimensional system with Euler class $\chi=2$ has four nodes, each carrying a winding number of $+1/2$ of the (real) eigenvector frame~\cite{Jiang_2021}.
\begin{figure}
\centering
\begingroup%
  \makeatletter%
  \providecommand\color[2][]{%
    \errmessage{(Inkscape) Color is used for the text in Inkscape, but the package 'color.sty' is not loaded}%
    \renewcommand\color[2][]{}%
  }%
  \providecommand\transparent[1]{%
    \errmessage{(Inkscape) Transparency is used (non-zero) for the text in Inkscape, but the package 'transparent.sty' is not loaded}%
    \renewcommand\transparent[1]{}%
  }%
  \providecommand\rotatebox[2]{#2}%
  \newcommand*\fsize{\dimexpr\f@size pt\relax}%
  \newcommand*\lineheight[1]{\fontsize{\fsize}{#1\fsize}\selectfont}%
  \ifx\svgwidth\undefined%
    \setlength{\unitlength}{242.99999567bp}%
    \ifx\svgscale\undefined%
      \relax%
    \else%
      \setlength{\unitlength}{\unitlength * \real{\svgscale}}%
    \fi%
  \else%
    \setlength{\unitlength}{\svgwidth}%
  \fi%
  \global\let\svgwidth\undefined%
  \global\let\svgscale\undefined%
  \makeatother%
  \begin{picture}(1,0.54962965)%
    \lineheight{1}%
    \setlength\tabcolsep{0pt}%
    \put(0,0){\includegraphics[width=\unitlength,page=1]{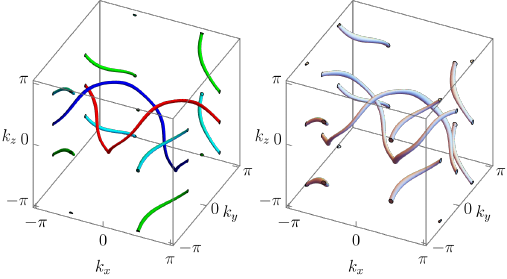}}%
    \put(0.01740187,0.49158995){\color[rgb]{0,0,0}\makebox(0,0)[lt]{\lineheight{1.25}\smash{\begin{tabular}[t]{l}\textbf{(a)}\end{tabular}}}}%
    \put(0.51122902,0.49158995){\color[rgb]{0,0,0}\makebox(0,0)[lt]{\lineheight{1.25}\smash{\begin{tabular}[t]{l}\textbf{(b)}\end{tabular}}}}%
  \end{picture}%
\endgroup%

  \caption{Correspondence between nodal lines and winding vector preimages in the balanced four-band Hopf-Euler insulator. \textbf{(a)} Preimages of the points $\pm\hat{\vb{x}}=(\pm 1, 0, 0)$ under the maps $\nh_\pm:T^3\to S^2$ from Sec. \ref{sec:4bandTB} with $m=1$ ($\nh^{-1}_+(\hat{\vb{x}})$ is shown in blue, $\nh^{-1}_+(-\hat{\vb{x}})$ in red, $\nh^{-1}_-(\hat{\vb{x}})$ in yellow and $\nh^{-1}_-(-\hat{\vb{x}})$ in cyan). Note that the helices are centered around the $\mathcal{C}_2$-invariant lines $\Gamma_y$ and $M_z$. \textbf{(b)} Nodal lines within the occupied subspace of the corresponding perturbed Hamiltonian from Eq. \eqref{eq:4BandPerturbed}.}
  \label{fig:4bandModels}
\end{figure}
\begin{figure}
\centering
   \def\svgwidth{.47\textwidth}
   \vspace{3em}
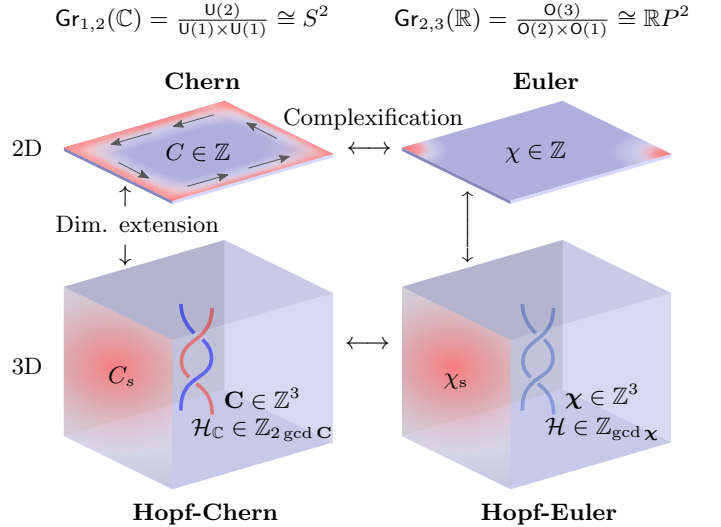
\vspace{1.3em}
   \caption{Dimensional extensions and complexification relations between the Chern, Euler, Hopf-Chern, and Hopf-Euler insulators; see also App.~\ref{app::G} for more details. We note that while the strong Hopf-Chern insulator $(\hhc; \textbf{C})=(1;0,0,0)$ displays surface states with surface Chern number $C_s = \hhc$, the Hopf-Euler insulator $(\hh; \boldsymbol{\chi})=(2;0,0,0)$ supports surface Euler states with a surface Euler invariant $\chi_s = \hh$; see App.~\ref{app::F}.}
 \label{fig_dim}
 \end{figure}

\subsection{Four-band models}\label{sec:4bandTB}
As shown in Sec. \ref{sec::II}, while the topological class of a real three-band model may be identified with that of a single winding vector $\dh(\vb{k})$, four-band models are classified by two winding vectors $\nh_\pm(\vb{k})$. It follows that models representing each four-band Hopf/Euler phase may be obtained by setting $\nh_\pm$ to be one of the winding vectors in Eqs. \eqref{eq:3bandModelStrongHopf}, \eqref{eq:3bandModelLayeredEuler} and \eqref{Eq:3BandExample}. Importantly, $\nh_-$ and $\nh_+$ can be chosen independently of one another.

We restrict our attention to a single example, namely a `balanced' Hopf-Euler insulator with winding vectors~\cite{bouhon2022multigap}
\begin{subequations}
\begin{align}
    \vb{n}_{-}(\vb{k}; m) ={}& \vb{d}_3(k_x, k_y, k_z; -m),\\
    \vb{n}_{+}(\vb{k}; m) ={}& \vb{d}_3(k_y, k_z, k_x; +m).    
\end{align}
\end{subequations}
The preimages of two points on $S^2$ under these maps are shown in Fig. \ref{fig:4bandModels}\textbf{(a)}. This system has topological invariants $\hh_\pm =\pm 1$, $\chi_{z-}=-2$, and $\chi_{y+}=2$, with all others vanishing. 

To move away from the degenerate flat-band limit, we add a perturbation $V_4 = \lambda \diag(-1, 0, 1, 0)$ to the Hamiltonian, giving
\begin{equation}\label{eq:4BandPerturbed}
    H_4(\vb{k}) = \bar{H}_4(\vb{k})+V_4.
\end{equation}
Here, without loss of generality, we again take $\lambda=0.8$ in all numerical studies. As with the three-band phases, the nodal lines within the occupied subspace of this four-band model lie in the same homotopy class as the preimages; see Fig. \ref{fig:4bandModels}\textbf{(b)}.
%

\section{Experimental realizations}\label{sec::V}
We now describe possible experimental realizations of the  three-dimensional real Hopf/Euler phases in metamaterials~\cite{RevModPhys.91.015006} and ultracold atoms, e.g.~by employing synthetic lattices~\cite{Cooper19_RevModPhs, Zhao_2022}. In particular, we suggest that an experiment of either type could in principle be designed to simulate Hopf-Euler insulators, and that in such an experiment it would be possible to observe quantum-geometric breathing and the presence of boundary modes with surface Euler invariants, and also to measure the bulk spectrum of the system along with the nodal helices that it displays.

In Sec.~\ref{sec::IV}, we gave a set of models that explicitly realize each of the phases discussed in this paper. As discussed at the end of Sec.~\ref{sec:3bandFlatModels}, the particular models representing the Hopf-Euler phase (see Eq. \eqref{Eq:3BandExample}) are constructed by `twisting' a two-dimensional Euler phase about the $k_z$ axis. This `dimensional extension' allows these models to be simulated experimentally by replacing this quasimomentum $k_z$ with a parameter $\lambda$, labeling a synthetic dimension. The parameter $\lambda$ could be, for instance, a label for a set of two-dimensional metamaterials~\cite{Jiang_2021, JIANG2024}, or a tunable parameter in a synthetic lattice~\cite{Cooper19_RevModPhs}. By measuring the two-dimensional system for a range of values of the parameter $\lambda$, the full three-dimensional spectrum of the model could be systematically reconstructed. This would then allow the physical properties of the system, such as the the quantum-geometric breathing of the Wannier functions, to be observed through wavefunction tomography experiments. 

\section{Discussion}\label{sec::VI}
We now discuss the presented theoretical results and physical manifestations of the Hopf-Euler insulators in the context of the electrodynamics of multi-gap topological phases. First, we note that the presence of quantum-geometric breathing explicitly demonstrates that there is no spectral flow present in these phases, that is, $\langle x \rangle = 0$ unlike the situation in Chern insulators (see App.~\ref{app::E},~\ref{app::G}, and Fig.~\ref{figPump}). In addition, since the $\theta$-angle $\theta = 2\pi \hh$, which follows from the definition of the invariant $\hh \in \mathbb{Z}$, there is no magnetoelectric effect present. This contrasts other known three-dimensional topological insulators such as the $\mathbb{Z}_2$ insulator with spinful time-reversal symmetry ($\mathcal{T}^2 = -1$, Altland-Zirnbauer class AII), and axion insulators with $\theta = \pi$~\cite{Qi_2008}. While the bulk of a Hopf-Euler insulator realizes a quantized optical effect, which by construction requires three spatial dimensions, this characteristic also sheds light on the electrodynamics of two-dimensional Euler insulators. In particular, the fact that this phase may be viewed as a dimensional extension of an Euler insulator suggests that no similar optical quantization is expected from Euler insulators. This is similar to two-dimensional phases not exhibiting quantized magnetoelectric effects, contrary to the three-dimensional bulks of topological insulators which realize $\theta$-vacua. As such, in two-dimensional Euler phases, any responses that can be captured with quantum geometry~\cite{provost1980riemannian, bouhon2023quantum,tormaessay}, are only manifested in terms of the lower bounds due to the topological invariants~\cite{PhysRevLett.124.167002, bouhon2023quantum}, rather than in terms of the quantization conditions, consistently with Ref.~\cite{jankowski2023optical}. In other words, the topology of Euler phases is purely quantum-geometric in its manifestations, contrary to the RHIs that support topological quantization in optical response through a bulk anomalous quantized circular shift effect on coupling to circularly polarized light~\cite{jankowski2024quantized}. This quantization might be deemed analogous to circular dichroism quantized by the Chern invariant in the Chern insulators. 
Indeed, while similarly to Hopf insulators, the multi-gap Euler phases are classified purely by homotopy theory ({e.g. $\pi_2[\mathbb{RP}^2] \cong \mathbb{Z}$}, in two spatial dimensions), the known bulk physical manifestations of Euler insulators consisted only of optical bounds~\cite{jankowski2023optical}, rather than quantized effects. This is contrary to the other two-dimensional insulators, such as Chern insulators, which support a bulk quantized quantum anomalous Hall effect, provided time-reversal-symmetry is broken~\cite{PhysRevLett.61.2015}.

We further comment on the procedure by which the Hopf-Euler phases in three dimensions were constructed, namely via dimensional extension~\cite{Qi_2008}, from two-dimensional Euler models. For a summary of the dimensional extension/reduction correspondences for complex and real Hopf insulators, see Fig.~\ref{fig_dim}. To recapitulate, the vector used for the construction of the Hopf-Euler Hamiltonian in Sec.~\ref{sec::IV} was constructed by extending the vector $\vb{d}(\kv)$, which  encodes an Euler invariant over a two-dimensional BZ~\cite{Unal_2020}, along an additional dimension $k_z$. This is achieved by multiplying $\vb{d}(\vb{k})$ by a $k_z$-dependent rotation matrix $R(k_z) = \ee^{\ii k_z L_z}$, which has the effect of twisting $\vb{d}(\vb{k})$ about the axis parallel to the $z$-direction, as shown in Sec.~\ref{sec::IV}. 
In particular, we reiterate that the rotation matrix $R(k_z)$ naturally promotes the nodes provided by the Poincar\'e-Hopf index theorem in two dimensions~\cite{PhysRevB.102.165151} to nodal helices, introduced in Sec.~\ref{sec::III}. Finally, it should be noted that Hopf-Chern insulators can, analgously, be obtained by extending two-band Chern Hamiltonians $H = \vec{d}(k_x,k_y) \cdot \boldsymbol{\sigma}$  with the same rotation, which must instead be taken to act on the winding vector $\vb{d}$ as $R(k_z) \vec{d}(k_x,k_y)$ Ref.~\cite{Kennedy_2016}. This construction therefore provides a natural connection between Hopf-Euler and Hopf-Chern insulators, which is induced by the complexification relations of the parent two-dimensional phases; see Fig.~\ref{fig_dim}. 

We finally remark that while four-band RHIs~\cite{Lim2023}, as well as the intrinsic bulk quantized shift currents that they display~\cite{jankowski2024quantized}, were studied in previous works; the strong three-band RHI introduced here was not included in previous topological classifications. We have thus extended the classification of $\mathcal{PT}$-symmetric phases not only to three-band RHIs, but also to Hopf-Euler insulators with weak homotopy invariants. These are naturally realizable by dimensional extensions of two-dimensional parent Euler Hamiltonians.

\section{Conclusions}\label{sec::VII}
We discuss a class of non-Abelian topological phases, namely three-band real Hopf insulators, with strong and weak homotopy invariants. We demonstrate that such Hopf insulators realize a topologically-quantized electromagnetic shift response, quantum-geometric breathing of real space hybrid Wannier functions, and boundary states with surface Euler invariant protected by $\mathcal{C}_2\mathcal{T}$ symmetry. We show these properties through analytical arguments and then give numerical demonstrations in minimal models. Our work offers a path for experimental realizations of such novel states of matter, as well as their physical signatures, such as nodal helices.
\begin{acknowledgements}
   W.J.J.~acknowledges funding from the Rod Smallwood Studentship at Trinity College, Cambridge. A.S.M.~acknowledges funding from EPSRC PhD studentship (Project reference 2606546).  A.B.~has been partly funded by a Marie Sklodowska-Curie fellowship, grant no. 101025315, and acknowledges financial support from the Swedish Research Council (Vetenskapsradet) (2021-04681). R.-J.S. acknowledges funding from a New Investigator Award, EPSRC grant EP/W00187X/1, a EPSRC ERC underwrite grant  EP/X025829/1, and a Royal Society exchange grant IES/R1/221060 as well as Trinity College, Cambridge.  F.N.\"U.~acknowledges funding from the Marie Sk{\l}odowska-Curie programme of the European Commission Grant No 893915, Simons Investigator Award [Grant No. 511029], and Trinity College Cambridge.  W.J.J. and R.-J.S. thank A. Alexandradinata for helpful discussions.
\end{acknowledgements}

\bibliography{references}

\newpage

\appendix

\section{The Pontryagin-Thom construction}\label{app::A}

\begin{figure}
\centering 
\def\svgwidth{.47\textwidth}
\vspace{1em}
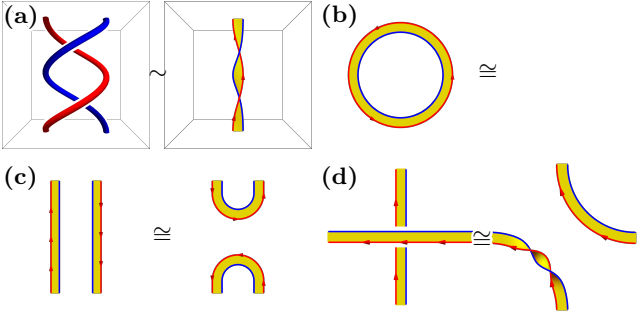
  \caption{The Pontryagin-Thom construction can be used to prove that the range of values of the strong invariant $\hh$ is reduced from $\mathbb{Z}$ to $\mathbb{Z}_{\gcd\boldsymbol{\chi}}$ in the presence of weak Euler invariants $\boldsymbol{\chi}$. \textbf{(a)}~The Pontryagin manifold ($\sim$ ``ribbon'') of a Hopf-Euler phase may be obtained by taking the preimage under $\ket{u_3}:T^3\to S^2$ of two infinitesimally separated points on $S^2$; here we show the case $(\hh; \boldsymbol{{\chi}}) = (1; 0, 0, 2)$. \textbf{(b-d)} Framed cobordisms of the preimage permitted by the topology-preserving transformations of the Hamiltonian. \textbf{(b)}~Cobordism of an empty set and a contractible loop without frame windings. \textbf{(c)}~Saddle cobordism. \textbf{(d)}~Undoing crossings via a cobordism introducing a frame winding.}
\label{Fig:knot}
\end{figure}

In this Appendix we will briefly describe the Pontryagin-Thom construction, which may be used to enumerate all the distinct topological classes of maps $T^3\to S^2$, as given in Eq. \eqref{eq:MapClasses}. For Hopf-Euler insulators, this map is given by the third eigenvector of the Bloch Hamiltonian. Our review closely follows Ref.~\cite{Kennedy_2016}, where the same construction was used to classify Hopf-Chern insulators. 

The Pontrjagin-Thom construction is used to generate a bijection between the set $[M, S^n]$, where $M$ is an $m$-dimensional oriented smooth manifold, and the set of framed cobordism classes of $(m-n)$-dimensional framed submanifolds of $M$, known as Pontrjagin manifolds. In the current context, the Pontrjagin manifold of a map $T^3\to S^2$ may be interpreted as the `ribbon' formed by the Seifert surface connecting two infinitesimally separated preimages on the sphere. For example, Fig.~\ref{Fig:knot}\textbf{(a)} shows the preimage of two points on $S^2$ in a Hopf-Euler phase with $(\hh; \boldsymbol{\chi})=(1; 0, 0, 2)$ on the left, and on the right the corresponding Pontrjagin manifold of the same phase; the latter may be obtained by taking an appropriate limit of the former. From here we see that the `twisting' of the ribbon corresponds to the Hopf invariant, while the fact that the ribbon connects the surfaces $k_z=\pm\pi$ shows that the phase has a non-zero Euler class. 

Also shown in Fig.~\ref{Fig:knot} are the maps that may be applied to a Pontrjagin manifold while leaving the topological class of the corresponding map $T^3\to S^2$ invariant (such maps are known as framed cobordisms, since they connect two framed submanifolds of $M$). By applying a particular combination of these homotopy operations to the Pontryagin manifold, such as shown in Fig.~\ref{Fig:knot}\textbf{(a)}, it is possible to remove the winding of the frame to obtain a ribbon with no twist. This shows that, in the presence of an Euler class, $\boldsymbol{\chi}=(0, 0, 2)$, e.g. the Hopf invariant $\hh=2$ is trivial, and may be removed with adiabatic transformations without closing the gap (i.e. $\hh$ is classified modulo $\gcd(\boldsymbol{\chi}) = 2$). We refer to Fig. 2 of Ref.~\cite{Kennedy_2016} for a precise description of the sequence of operations required in this case.


\section{Quaternions and minimal models}\label{app::B}\label{app:Quaternion}

In this Appendix, we describe the quaternion formulation~\cite{Unal_2020,Lim2023} of the real Hopf invariant, and relate it to three-dimensional non-Abelian topological insulators. 

\subsection{Quaternions and rotations}
We begin with a short review of quaternions, before using them to provide a concise formulation of the Hopf map. The quaternion algebra $\mathbb{H}$ is defined to be the real vector space with basis vectors $\{1, \ii, \jj, \kk\}$ where $\ii^2=\jj^2=\kk^2 = \ii\jj\kk = -1$, that is, $\mathbb{H} = \left\{a + b\ii +c\jj +d\kk\,\middle|\, a, b, c, d \in \mathbb{R}\right\}$. In the following we will let $p=p_0 + p_1\ii +p_2\jj +p_3\kk$ and $q=q_0 + q_1\ii +q_2\jj +q_3\kk$ be elements of $\mathbb{H}$, which are known as quaternions. The real and imaginary parts of $q$ are respectively defined as $\Re[q]=q_0$ and $\Im[q] =q_1\ii +q_2\jj +q_3\kk$. The conjugate $\bar{q}$ of $q$ is given by reversing the sign of the imaginary part of $q$, that is, $\bar{q} = q_0 - q_1\ii -q_2\jj -q_3\kk$. It is useful to note that $\bar{q}=-\frac{1}{2}(q+\ii q\ii+\jj q\jj+\kk q\kk)$, and also that the conjugate of a product of two quaternions $p$ and $q$ satisfies $\overline{p q}=\bar{q}\bar{p}$ (note that the order of multiplication is reversed). A real quaternion is one for which $\Re[q]=q$ or, equivalently, $\bar{q}=q$, and simiarly an imaginary quaternion has $\Im[q]=q$ and $\bar{q}=-q$. The norm of $q$ is $|q|=\sqrt{q\bar{q}} = \sqrt{q_0^2+q_1^2+q_2^2+q_3^2}$, and satisfies $|\bar{q}|=|q|$ and $|p q|=|p||q|$. A quaternion of unit norm (i.e. for which $|q|=1$) is known as a versor. Finally, we note that the Euclidean inner product $\braket{p}{q}=\sum_{\mu=0}^3 p_\mu q_\mu$ may be written in terms of these quaternionic operations as $\braket{p}{q} = \Re[\bar{p}q]=\Re[p\bar{q}]$.

Within the algebra $\mathbb{H}$, the set of quaternions with unit norm (versors) is special, since it is isomorphic as a group to $S^3\cong\mathsf{SU}(2)\overset{\pi}{\rightarrow}\mathsf{SO}(3)$. As is well known, this provides a means by which rotations in three dimensions can be described in terms of quaternions. To be specific, by identifying $\mathbb{R}^3$ with the set of imaginary quaternions $\mathbb{H}^*$ in $\mathbb{H}$, the action of a versor $v=x_0+x_1\ii +x_2\jj +x_3\kk$ on a vector $\vb{u}=(u_1, u_2, u_3)$, or imaginary quaternion $u=u_1\ii+u_2\jj+u_3\kk$, given by
\begin{equation}\label{eq::versorrot}
R_v: u \mapsto \bar{v} u v,    
\end{equation}
implements a rotation by an angle $\theta=2\arccos{(x_0)}=2\arcsin{(1-x_0)}$ around the vector $\mathbf{v}=(x_1, x_2, x_3)$. The corresponding $\mathsf{SO}(3)$ matrix $R_v$ may be written down by acting with the versor $v$ on each of the unit vectors $\ii$, $\jj$ and $\kk$, and collecting the results as column vectors:
\begin{equation}
    R_v = \mqty(\ket{\bar{v}\ii v} && \ket{\bar{v}\jj v} && \ket{\bar{v}\kk v}),
\end{equation}
which is used in Eq. \eqref{eq:SO3Rot}. For completeness, we give the explicit form of this matrix in terms of the parameters $x_\mu$ in $v$~\cite{Unal_2020}, 
\begin{widetext}
\begin{equation}\label{eq:RotMat3}
    R_v = \mqty(x_0^2+x_1^2-x_2^2-x_3^2 & 2(x_1 x_2 +x_0 x_3) & 2(x_1 x_3 -x_0 x_2) \\ 2(x_1 x_2-x_0 x_3) & x_0^2-x_1^2+x_2^2-x_3^2 & 2(x_0 x_1+ x_2 x_3) \\ 2(x_0 x_2+x_1 x_3) & 2(-x_0 x_1+x_2 x_3) & x_0^2-x_1^2-x_2^2+x_3^2). 
\end{equation}
\end{widetext}

\subsection{Three-band models and the Hopf invariant}\label{app::Quaternions.3band}
The form of Eq. \eqref{eq:RotMat3} may be understood by noting that, for any versor $v$, each of the elements $\bar{v}\ii v$, $\bar{v}\jj v$ and $\bar{v}\kk v$ is a purely imaginary quaternion of unit norm. This means that they are elements of the set $\{q_1\ii+q_2\jj+q_3\kk\,|\, q_1^2+q_2^2+q_3^2=1\}$, in other words, they lie on the sphere $S^2$. This shows that the parametrization of the rotation matrix given above directly induces, for each versor, a map from $S^3$ to $S^2$ transversely. In this way, the matrix in Eq. \eqref{eq:RotMat3} presents a general parametrization of the Hopf map, in the sense that any row, column or linear  combination (of unit norm) from the left or right implements the first Hopf map. This provides a means by which the  three-band models of interest may be easily formulated.

Finally, using the quaternion formulation described above, we derive the relation between the Euler connection and the one-form present in the Whitehead formula. Using the definitions provided in the main text, one may deduce that 
\begin{align}
    \begin{split}
        \aa = \text{Pf}[-\ii \mathrm{A}] ={}& \braket{u_1}{\dd u_2} \\
        ={}& \bra{\bar{q}\kk q}\dd\ket{\bar{q}\jj q} \\
        ={}& \Re[\bar{q}\kk q \dd(-\bar{q}\jj q)]\\
        ={}& -\Re[\bar{q}\kk q(\dd\bar{q}\cdot \kk z+\bar{q}\kk\dd q) ]\\
        ={}& -\Re[\jj q\dd\bar{q}\cdot\kk]-\Re[\bar{q}\jj \kk \dd q]\\
        ={}& -2\Re[\ii q\dd\bar{q}]\\
        ={}& 2\omega,
    \end{split}
\end{align}
where we have used $|q|^2=1$. This concludes the proof of the identity $\aa = 2\omega$ utilized in the main text.

\subsection{Four-band models}\label{app::Quaternions.4band}

As noted in Eq.~\eqref{eq:4BandFlat} in the main text, a flattened four-band real Hamiltonian may be written in either of the equivalent forms
\begin{align}
\begin{split}
    \bar{H}_4(\vb{k}) ={}& \nh_+(\kv)\cdot \underline{\Gamma} \cdot \nh_-(\kv)\\
    ={}& R_4(\vb{k})\diag\mqty(1, & 1, & -1, & -1) R_4(\vb{k})^\text{T},
\end{split}
\end{align}
where $\underline{\Gamma}$ is an array of Dirac matrices and $R_4(\vb{k})$ is an $\mathsf{SO}(4)$ matrix \cite{bouhon2022multigap}. We firstly note that, due to the isormorphism $\mathsf{SO}(4)\cong[\mathsf{SU}(2)\times\mathsf{SU}(2)] / \mathbb{Z}_2$, the matrix $R_4$ may be written in terms of \textit{two} quaternions $q_\pm$,
\begin{align}
    R_4 ={}& \mqty(\ket{\overline{q}_- q_+} && \ket{\overline{q}_-\ii q_+} && \ket{\overline{q}_- \jj q_+} && \ket{\overline{q}_- \kk q_+}).
\end{align}
This is discussed extensively in Sec.~\ref{sec:FourBandPhases}, as well as in Ref.~\cite{Lim2023}, where it is used to classify four-band real Hopf insulators.

For completeness, we also give the explicit form of $\underline{\Gamma}$ in terms of a set of Dirac gamma matrices $\Gamma_{\mu\nu} = \sigma_\mu \otimes \sigma_\nu$, where $\mu, \nu=0, 1, 2, 3$:
\begin{equation}\label{eq:GammaArray}
\underline{\Gamma} =
\mqty(\Gamma_{30} & \Gamma_{22} & \Gamma_{10} \\
    \Gamma_{11} & \Gamma_{03} & -\Gamma_{31} \\
    -\Gamma_{13} & \Gamma_{01} & \Gamma_{33}).
\end{equation}
All matrices within the array $\underline{\Gamma}$ are real, so that the resulting Hamiltonian $\bar{H}_4(\vb{k})$ manifestly respects $\mathcal{PT}$ symmetry. We note that this array differs from that employed in \cite{bouhon2022multigap}, which is given by
\begin{equation}\label{eq:GammaArrayTilde}
\underline{\tilde{\Gamma}} =
\mqty(-\Gamma_{33} & -\Gamma_{13} & \Gamma_{01} \\
    \Gamma_{31} & \Gamma_{11} & \Gamma_{03} \\
    \Gamma_{10} & -\Gamma_{30} & -\Gamma_{22}).
\end{equation}
It may be verified that the conventions laid out in Eqs. \eqref{eq:GammaArray} and \eqref{eq:GammaArrayTilde} are simply related by a change of basis,
\begin{equation}
    \underline{\Gamma}_{\alpha\beta} = \sum_{\gamma,\delta=1}^3 M^{+}_{\gamma\alpha}\underline{\tilde{\Gamma}}_{\gamma\delta} M^{-}_{\delta\beta},
\end{equation}
where the two matrices $M^{\pm}$ are given by
\begin{align}
    M^{+} = \mqty(0 & 0 & 1 \\ 0 & 1 & 0 \\ -1 & 0 & 0), \qquad M^{-}=\mqty(0 & 0 & -1\\ 1 & 0 & 0 \\ 0 & 1 & 0).
\end{align}

\section{Four-band flag phases}\label{app::C}
We now complete our discussion of four-band real phases by considering the effect that opening a gap within the valence or conduction bands has on the topology of the system. As in Sec.~\ref{sec:3bandflag}, we approach this question by classifying the possible flag manifolds that may arise in each case.

Firstly, we consider the case where the system is fully gapped and no two bands touch at any point in the BZ. The classifying space for these four-band real flag phases is $\mathsf{Fl}_{1, 1, 1, 1}=\mathsf{O}(4)/[\mathsf{O}(1)^4]$~\cite{PhysRevB.109.165125}, and the relevant homotopy groups of the classifying space are
\begin{equation}
    \pi_k[\mathsf{Fl}_{1, 1, 1, 1}] \cong\pi_k[\mathsf{SO}(4)]\cong\pi_k[S^3]\times\pi_k[S^3]\cong \mathbb{Z}^2,
\end{equation}
where $k=2, 3$, and we have made use of the isomorphisms $\mathsf{SO}(4)\cong\mathsf{SU}(2)\cross\mathsf{SU}(2)/\mathbb{Z}_2$ and $\mathsf{SU}(2)\cong S^3$. The group $\pi_2[S^3]$ is trivial while, as discussed in Sec.~\ref{sec:3bandflag}, $\pi_3[S^3]\cong \mathbb{Z}$ is labeled by a Pontryagin index. We therefore see that the fully-gapped flag phase can be labeled by \textit{two} Pontryagin indices $w_\pm$. We now describe how an explicit expression for each of these invariants may be obtained. The representative Hamiltonians of the four-band flag phases can be constructed analogously to the three-band phases as
\begin{equation}
    \bar{H}^{\text{flag}}_4 = V_4(\vb{k})\diag\mqty(2, &  1, &  -1, & -2) V_4(\vb{k})^\text{T},
\end{equation}
where $V(\vb{k})\in\mathsf{SO}(4)$. This matrix may be written in the form $V(\vb{k}) = \exp(\ii\boldsymbol{\theta}_+(\vb{k})\cdot\vb{L}_++\ii\boldsymbol{\theta}_-(\vb{k})\cdot\vb{L}_-)$, where ${\vb{L}_\pm=(\vb{J} \pm \vb{K}})/2$ with
\begin{subequations}
    \begin{align}
        (K_i)_{\mu\nu} ={}& \ii(\delta_{0, \mu}\delta_{i \nu}-\delta_{0, \nu}\delta_{i\mu})\\
        (J_i)_{\mu\nu} ={}&\frac{\ii}{2}\epsilon_{ijk}(\delta_{k\mu}\delta_{j\nu}-\delta_{k\mu}\delta_{j\mu})
    \end{align}
\end{subequations}
and the indices $i = 1,2,3$ and $\mu, \nu=0, 1, 2, 3$. The $\vb{L}_\pm$ generate the $\mathfrak{so}(4)$ Lie algebra, and moreover we have chosen a basis in which each set $\vb{L}_+$ and $\vb{L}_-$ generates one of the two $\mathfrak{su}(2)$ algebras which together make up this space. In this way, by using the parameters $\boldsymbol{\theta}_\pm$ we can produce two $\mathsf{SU}(2)$ matrices
\beq{}
    U_\pm(\vb{k}) = \exp(\ii\boldsymbol{\theta}_\pm(\vb{k})\cdot\boldsymbol{\sigma}/2).
\eeq    
which may be used to compute each of the Pontryagin indices $w_\pm\in\mathbb{Z}$. We have~\cite{Zee2010-ZEEQFT, PhysRevB.109.165125},
\begin{align}
\begin{split}
    w_{\pm} ={}& \frac{1}{24\pi^2} \intBZ \text{Tr} [(U^{-1}_{\pm}\dd U_{\pm})^3]\\
    ={}& -\frac{1}{16\pi^2}\intBZ (\aa^{\text{c}}\mp\aa^{\text{v}})\wedge(\Eu^{\text{c}}\mp\Eu^{\text{v}}).
\end{split}
\end{align}
Similarly to the three-band flag phases discussed in Sec.~\ref{sec:3bandflag}, while the right hand side of this expression is exactly the same as that used to compute the Hopf invariants in Eq. \eqref{eq:HopfPM} (see also~\cite{PhysRevB.109.165125}), in this case it instead computes the Pontryagin indices of the system. 

In addition to requiring all bands to be open, we can individually close a gap between two lowest, or two highest, energy bands. Since these are both described by $\mathsf{Fl}_{2, 1, 1}\cong\mathsf{Fl}_{1, 1, 2}$, respectively, we use
\begin{equation}
    \pi_k[\mathsf{Fl}_{2, 1, 1}]\cong\pi_k\left[\frac{\mathsf{O}(4)}{\mathsf{O}(2)\times\mathsf{O}(1)^2}\right]\cong\pi_k[S^2]\times\pi_k[S^3]
\end{equation}
where we have noted that $\mathsf{SO}(4)/\mathsf{SO}(2)\cong S^2\times S^3$. A~natural interpretation of this correspondence is that opening a gap causes the change $S^2\rightsquigarrow S^3$ in one of the factors of the Grassmannian classifying the four-band real Hopf insulator~\cite{Lim2023}. Another way to view this is as follows. Starting from the fully-gapped flag limit $\mathsf{Fl}_{1, 1, 1, 1}\sim S^3\times S^3$, bringing two bands together introduces a gauge degree of freedom that introduces a factor of $\mathsf{SO}(2)$. Since $S^3/\mathsf{SO}(2)\sim \mathsf{SO}(3)/\mathsf{SO}(2) \cong S^2$, this has the effect of reducing the dimension of one of the spheres in the classifying space (here we have used $\sim$ to denote congruence modulo discrete factors). Thus, while we saw above that opening both gaps causes all Euler invariants to become trivial and both Hopf invariants to become Pontryagin invariants, it is clear that opening a single gap trivializes a \textit{single} set of Euler invariants (either $\boldsymbol{\chi}_+$ or $\boldsymbol{\chi}_-$), and causes \textit{one} Hopf invariant to become a Pontryagin invariant (the other remains a Hopf invariant).

%
Finally, having closed the first gap, if a third band is joined on to this group of two, then the classifying space becomes $\mathsf{Gr}_{1, 4}(\mathbb{R})=\mathsf{O}(4)/[\mathsf{O}(3)\times\mathsf{O}(1)]\sim \mathsf{SO}(4)/\mathsf{SO}(3)\cong S^3$, so that the system is characterized by a single (strong) Pontryagin index~\cite{Zee2010-ZEEQFT,PhysRevB.109.165125} and no Euler classes. This case was discussed in detail in  Ref.~\cite{PhysRevB.109.165125}.

\section{Quantized circular shift effect}\label{app::D}
We now derive the symmetrized quantized shift effect in the 3-band RHI models enjoying the reality condition due to the $\mathcal{PT}$ symmetry, following Ref.~\cite{jankowski2024quantized}. First, we define the torsion tensor $T^{mn}_{ijk} \equiv C^{mn}_{ijk} - C^{mn}_{ikj}$ with the Hermitian connection $C^{mn}_{ijk}$ defined in the main text. With little algebra, it can be directly shown that the torsion tensor may be written in terms of the non-Abelian Berry connection as~\cite{jankowski2024quantized}
\begin{subequations}
\begin{align}
     T^{31}_{ijk} + T^{31}_{jki} + T^{31}_{kij} ={}& 
    A^{[i}_{13} A^k_{32} A^{j]}_{21},\\
    T^{32}_{ijk} + T^{32}_{jki} + T^{32}_{kij} ={}& 
    A^{[i}_{23} A^k_{31} A^{j]}_{12}.   
\end{align}
\end{subequations}
where $[\ldots]$ denotes antisymmetrization of the enclosed indices $i,j,k$. On summing all the terms and using $A^i_{nm}=-A^i_{mn}$, which is enforced by the $\mathcal{PT}$ symmetry, we obtain
\begin{align}
    & \sum^{\text{occ}}_{n} \sum^{\text{unocc}}_{m} (T^{mn}_{ijk} + T^{mn}_{jki} + T^{mn}_{kij}) = \\& -2A^i_{12}A^{[j}_{13}A^{k]}_{32}
     -2A^j_{12}A^{[k}_{13}A^{i]}_{32}
    -2A^k_{12}A^{[i}_{13}A^{j]}_{32}.\nonumber
\end{align}
where `\text{occ}' and `\text{unocc}' denote the occupied and unoccupied bands, respectively.~Furthermore, one can define a two-band Euler curvature: $\text{Eu}^{ij}_{nm} = \bra{\partial_i u_{[n}}\ket{\partial_j u_{m]}}= \sum_{p \neq n, m} A^{[i}_{np} A^{j]}_{pm}$, to simplify the expression as,
\beq{}
\begin{split}
    \sum_{n, m} (T^{mn}_{ijk} + T^{mn}_{jki} + T^{mn}_{kij}) = -2A^{(i}_{12} \text{Eu}^{jk)}_{12}.
\end{split}
\eeq
where $(\ldots)$ denotes a symmetric sum over permutations in the indices $i,j,k$.
Here, the Euler connection ${\textbf{a} = \vec{A}_{21} = -\vec{A}_{12}}$, and the elements of the Euler (pseudo)vector (see the main text) can be identified as $(\text{\textbf{Eu}})_i = \frac{1}{2} \epsilon_{ijk} \text{Eu}^{jk}_{12}$. We then find
\beq{}
    \sum_{n,m} (T^{mn}_{ijk} + T^{mn}_{jki} + T^{mn}_{kij}) = -2~\vec{a} \cdot \vb{Eu}.
\eeq
Upon substituting the Hermitian connection in terms of the torsion tensor and Euler curvatures/connections into the shift photoconductivity formula~Eq.~\eqref{eq:shift}, we obtain the part of the shift photoconductivity that couples to the \textit{circularly polarized} light, $\sigma^{ijk}_{\text{shift,C}} = \mathfrak{Im}~\sigma^{ijk}_{\text{shift}}$~\cite{PhysRevX.10.041041},
\begin{align}
\begin{split}
    F_{\text{sym}} ={}& \int \dd \om~ \Big[\sigma^{ijk}_{\text{shift,C}}(\om) + \sigma^{jki}_{\text{shift,C}}(\om) + \sigma^{kij}_{\text{shift,C}}(\om)\Big] \\
    ={}& -\frac{e^3}{8\pi^2} \int_{\text{BZ}}  \dd^3\kv \sum_{m,n} f_{nm} (T^{mn}_{ijk} + T^{mn}_{jki} + T^{mn}_{kij}) \\
    ={}& \frac{e^3}{4\pi^2} \int_{\text{BZ}} \dd^3\kv~
    \vb{a} \cdot \vb{Eu} =  \frac{2e^3 }{\hbar^2}\hh,
\end{split}
\end{align}
where in the last equality we have identified the Hopf invariant $\hh$ via Eq. \eqref{eq:HopfIntegral} and restored the reduced Planck constant $\hbar$.

It should be noted that for the existence of the second-order quantized shift response on the left-hand side, inversion symmetry $\mathcal{P}$ must be broken. Indeed if the $\mathcal{P}$ symmetry is preserved, then $\hh=0$, which follows from the transformation of the Euler (pseudo)vector and the Euler connection under this symmetry. More precisely, inversion symmetry enforces the constraint 
\beq{}
    \textbf{Eu}(\kv) = \textbf{Eu}(-\kv),
\eeq
for the Euler (pseudo)vector, and 
\beq{}
    \textbf{a}(\kv) = -\textbf{a}(-\kv).
\eeq
for the Euler connection. Therefore, the integrand of the real Hopf invariant is odd under inversion, so the integral of this quantity over the entire $\text{BZ} \cong T^3$ vanishes. In the four-band case, it was shown analogously that~\cite{jankowski2024quantized}
\beq{eq:FHopf}
\begin{split}
    F_{\text{sym}} = \frac{2e^3}{\hbar^2} (\hh_{+} + \hh_{-});
\end{split} 
\eeq
inversion symmetry forces $\hh_{-} + \hh_{+} = 0$~\cite{Lim2023}, again suppressing the second-order response.

\section{Quantum metric bounds and breathing}\label{app::E}
In this section we outline the relation between the quantum metric and the Euler class, in terms of quantum geometric bounds. Moreover, as outlined in the main text, we retrieve the quantum-geometric breathing in RHIs, both analytically and numerically.

The quantum metric is most easily expressed in terms of the projector $\hat{P}$ into the occupied bands, which is given by $\hat{P} = \sum_a^{\text{occ}} \ketbra{u^a(\vb{k})}{u^a(\vb{k})}$, with $\kv=(k_x,k_y,k_z$) and `\text{occ}' the occupied bands,
\begin{equation}
    g_{ij} \equiv \frac{1}{2}\text{Tr}[(\partial_i \hat{P})(\partial_j \hat{P})].
\end{equation}
As mentioned in the main text, the quantum metric gives a bound on the second-moment/variance of the Wannier functions through the Resta-Sorella relation~\cite{PhysRevLett.82.370}, which underlies QGB, and which we define next.

Namely, to demonstrate QGB, we first obtain Wannier functions~\cite{RevModPhys.84.1419}, for a general case of a $d$-dimensional system, by taking the inverse Fourier transform of Bloch states ${\ket{\psi^a(\vb{k})} = e^{\ii \kv \cdot \hat{\vec{r}}} \ket{u^a(\vb{k})}}$,
\beq{}
    \ket{w_{n\R}} = \intBZ \frac{\dd^d \kv}{(2\pi)^d}~e^{-\ii \kv \cdot \vec{R}} \ket{\psi^a(\vb{k})},
\eeq
where $\ket{w_{n\R}}$ is a Wannier state in unit cell at a position vector $\R$. We rescale the $d$-dimensional volume of a unit cell to unity, $V_d = 1$, here, and in the subsequent section. Moreover, a Wannier center in a cell at the position $\R$ is defined as,
\beq{}
    \bar{w}_{n\R} = \bra{w_{n\R}} \hat{\vec{r}} \ket{w_{n\R}}. 
\eeq
Additionally, we define $hybrid$ Wannier states by a Fourier transform in a single direction, without loss of generality $x$,
\beq{}
    \ket{w^x_{n\R}} = \intBZ \frac{\dd k_x}{(2\pi)}~e^{-\ii k_x \cdot (\vec{R})_x} \ket{\psi^a(\vb{k})},
\eeq
and the hybrid Wannier centers read
\beq{}
    \bar{w}^x_{n\R}(k_y, k_z) = \bra{w^x_{n\R}} \hat{x} \ket{w^x_{n\R}}.
\eeq
It should be noted that the real-space basis used to express the Wannier functions is a basis of states at localized positions $(0,\ldots,0, 1,0,\ldots,0)$, resembling a basis of Dirac $\delta$-functions in a lattice-regularized realization, cf. in a continuum limit ${\ket{\textbf{r}}}$, constitutes a complete basis in the position representation, with $\bra{\textbf{r}'}\ket{\textbf{r}} = \delta(\textbf{r}-\textbf{r}')$. In other words, the Wannier functions in the three-dimensional systems central to this work, are defined as $w_{n\R}(\rv) = \bra{\textbf{r}}\ket{w_{n\R}}$, whereas the hybrid Wannier functions (HWFs) are given by $w^x_{n\R}(\textbf{r}, k_y, k_z) = \bra{\textbf{r}}\ket{w^x_{n\R}}$. In both cases, we only consider $\R = \vec{0}$, and the moduli square of the Wannier functions; $|w|^2 \equiv |w_{n\vec{0}}(\rv)|^2$ and $|w_x|^2 \equiv |w^x_{n\vec{0}}(\textbf{r}, k_y, k_z)|^2$, correspondingly. 

Now, with all the definitions related to the Wannier functions and relevant quantum geometry at hand; we focus on the Euler insulators defined in the main text. First, we note that for any two-band Euler subspace $\{ \ket{u_1}, \ket{u_2} \}$, we can rewrite the bands in a complexified, Chern basis,
\beq{}
\ket{u^{\pm}} = \frac{1}{\sqrt{2}}(\ket{u_1} \pm \ii \ket{u_2}),
\eeq
which, when Fourier-transformed, yields the corresponding Wannier orbitals $\ket{w^{\pm}_{\R}}$ with opposite quantized charge flows akin to Thouless pumping~\cite{PhysRevB.27.6083}. Namely, the complexified charge centers $\bar{w}^{\pm}_\R$ acquire equal and opposite Berry phases $\phi_x$~\cite{vanderbilt2018berry}, consistently with the non-Abelian Wilson loop winding of the Euler insulators. In other words, the complexified centers,
\\
\beq{}
\bar{w}^{\pm}_\R = \bra{w^{\pm}_\R} \hat{x} \ket{w^{\pm}_\R}
\eec
\\
wind by $2\pi\chi$ on a full parameter $k_z$ cycle. This follows from the fact that on direct evaluation of the invariants with complexified bands, $\chi = \frac{C_+ - C_-}{2}$, with $C_\pm$ representing the Chern numbers of the complexified bands~\cite{bouhon2018wilson, Bouhon_2019, JIANG2024}. Indeed, this needs to be definitionally true from the curvature-based definitions of corresponding characteristic classes, given the bundle complexification ($E \oplus iE$) relation between the Euler and Chern characteristic classes, namely, $\chi(E) = C(E \oplus iE)$, where $E$ denotes a vector bundle~\cite{Nakahara, Bouhon_2019, JIANG2024}.
As the winding of the complexified Wannier orbital centers $w^{\pm}_\R$, can be rewritten in terms of the original Wannier orbitals as a winding of $\bra{w_1} \hat{x} \ket{w_2}$, this shows that such a \textit{complexified} Thouless pump of Euler insulators (see also App.~\ref{app::G}, and Fig.~\ref{figPump}) can be viewed as a flow of interband dipoles between bands, apart from the complexified (Chern basis) picture of a superposition of two counter-propagating charge pumps, as captured by the Berry phases.

\begin{figure*}
\centering
  \includegraphics[width=2.0\columnwidth]{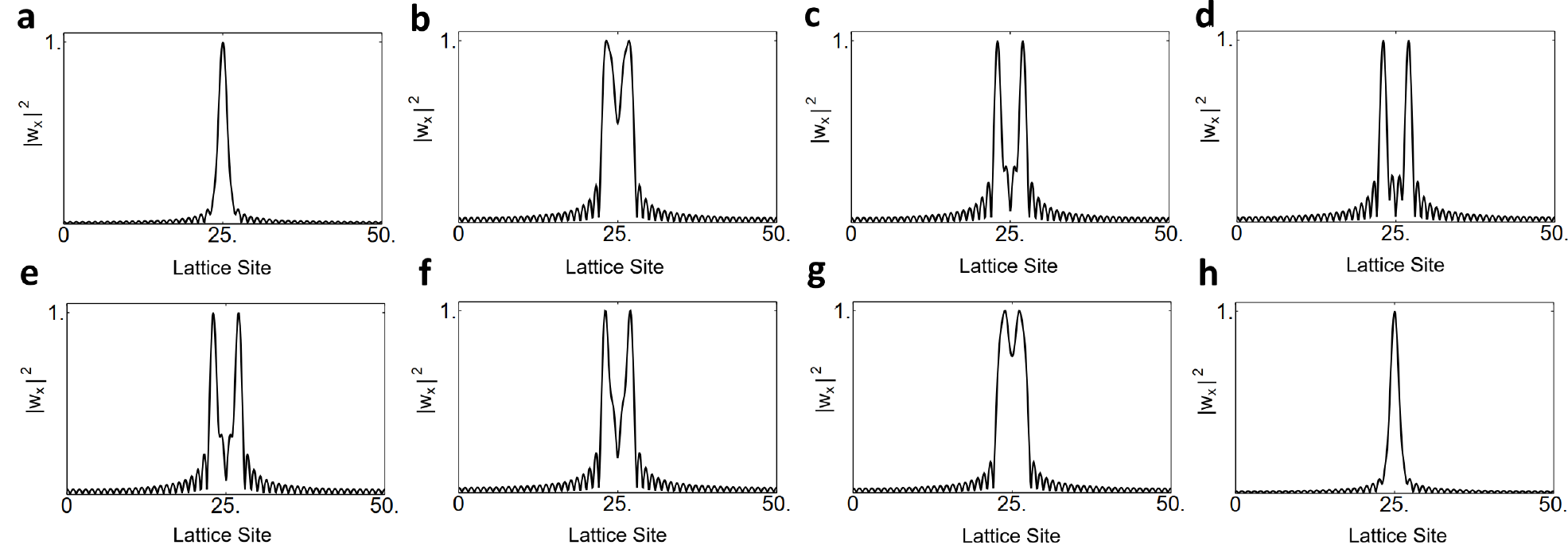}
  \caption{Quantum-geometric breathing in the \textit{three-band} RHIs. The variance of the RHI hybrid Wannier functions changes as the parameter $k_z$ evolves. Multiple sections correspond to different breathing stages of the hybrid Wannier localized at a fixed lattice site. (\textbf{a}--\textbf{h}). The modulus square of the maximally-localized Wannier function at parameters $k_z = [0,~2.7,~3,~3.2,~3.3,~3.4,~3.7,~6.2]$. Here, the three-band models with the strong real Hopf index $\hh=2$, and weak Euler indices, realize a dimensional extension of the three-band Euler insulator.}
\label{figQGB3}
\end{figure*}

\begin{figure*}
\centering
  \includegraphics[width=2.0\columnwidth]{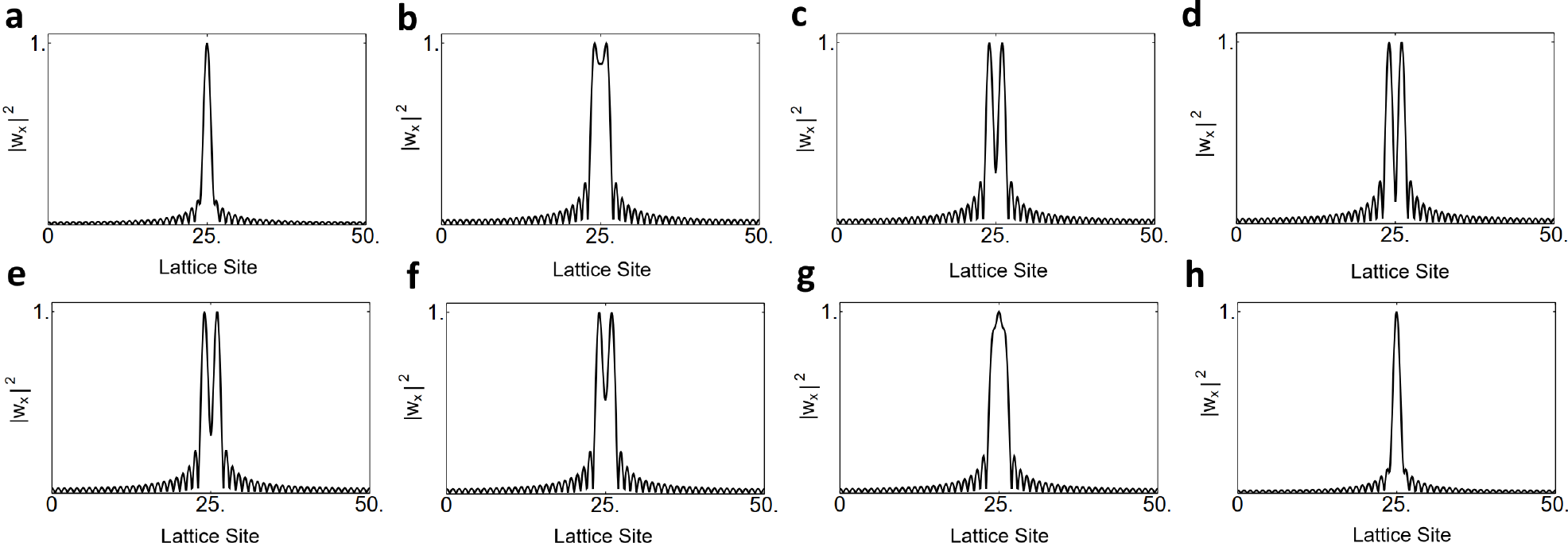}
  \caption{Quantum-geometric breathing in \text{four-band} RHIs. As the parameter $k_z$ evolves, the variance of the RHI hybrid Wannier functions changes. Multiple sections correspond to different breathing stages of the hybrid Wannier localized at a fixed lattice site. (\textbf{a}--\textbf{h}). The modulus square of the Wannier function at parameters $k_z = [0,~2.7,~3,~3.2,~3.2,~3.3,~3.4,~3.7,~6.2]$. Here, the four-band models with the strong real Hopf indices $(\hh_+,\hh_-)=(0,1)$ and weak indices present realize a dimensional extension of the four-band Euler insulators.}
\label{figQGB4}
\end{figure*}

In a higher-dimensional context, with a completely analogous reasoning based on the complexification of the bands $\ket{u^\pm}$, or equivalently, Wannier states $\ket{w_{n\textbf{R}}}$, the returning Thouless pump (RTP) of a Hopf insulator~\cite{alexandradinata2021,Nelson_2022} can be complexified, and hence transferred/realized in an occupied band subspace of a three-band RHI. The RTP amounts to pumping a Wannier center by $\hhc$ unit cells halfway through a pumping cycle, and subsequently restoring the original position of the Wannier center on the return in a full pumping cycle~\cite{alexandradinata2021,Nelson_2022}. Intuitively, a complexified pair of such opposite RTPs (see Fig.~\ref{figPump}), results in the quantum-geometric breathing derived below, which is furthermore demonstrated numerically. Namely, the $spread$ of the Wannier functions oscillates to the extent of $|\hh|$ unit cells (or $|\hh_+ + \hh_-|$ in a four-band case), as two complexified centers perform a full RTP cycle in the RHIs. Accordingly, the maximal spread is obtained halfway through the cycle $k_z = \pi$, which corresponds to the saturation of a quantum-geometric ``breathe", at the peak displacement of the complexified centers. For the numerical demonstration of the described quantum-geometric breathing (QGB) in the three-band and four-band Hamiltonians, see Figs.~\ref{figQGB3},~\ref{figQGB4}.

Before continuing to analytically retrieve the QGB directly from the introduced Wannier functions and quantum metric itself, we utilize the desribed complexified band picture to derive a bound on the quantum metric, due to the Euler curvature. To achieve that, we note that, a matrix
\beq{}
\tilde{Q}^+_{ij} = \bra{\partial_{k_i}u^+}(1-\hat{P})\ket{\partial_{k_j}u^+},
\eeq
with $i,j$ taking only two out of three values $x,y,z$, is by construction positive-semidefinite~\cite{bouhon2023quantum,jankowski2023optical}. Therefore, $\text{Tr}~{Q}^+_{ij} \geq 0$. Then, upon rewriting the complexified bands in terms of the original bands $\ket{u_1}, \ket{u_2}$, with Euler curvature defined in Sec.~\ref{sec::II}, one retrieves 
\beq{}
    \text{Tr}~ g_{ij} - 2 \text{Eu}_{ij} \geq 0.
\eeq
Analogously, on repeating the steps with similarly-defined $\tilde{Q}^-_{ij}$, in terms of $\ket{u^-}$, one obtains,
\beq{}
\text{Tr}~g_{ij} + 2 \text{Eu}_{ij} \geq 0.
\eeq
Combining two inequalities yields the final result, used in the main text,
\beq{}
    g_{ii} + g_{jj} \geq 2 |{\text{Eu}_{ij}}|,
\eeq
where we stress that $i,j$ take only two out of three possible values, under the inequality.
 
We now move to the final steps of an explicit derivation of QGB. As due to the $\mathcal{PT}$ symmetry $\langle x \rangle = 0$, we address the dependence of the variance ${\sigma^2_r(k_z) = \langle x^2 \rangle - \langle x \rangle^2 = \langle x^2 \rangle}$ on $k_z$, definitional for QGB. Namely, we recognize that on hybrid-Wannierizing occupied bands $\ket{u_{1,2}}$, by Fourier transforming in $x$, which obtains hybrid Wannier states $\ket{w^x_{1,2}}$, we have,
\begin{align}
    \begin{split}
        \langle x^2 \rangle ={}& \bra{w^x_{1,2}} x^2 \ket{w^x_{1,2}} (k_z)\\
        ={}& \bra{w^x_+} x^2 \ket{w^x_+} (k_z) + \bra{w^x_-} x^2 \ket{w^x_-} (k_z)\\
        \geq{}& (|\bra{w^x_+} x \ket{w^x_+}|^2 + |\bra{w^x_-} x \ket{w^x_-}|^2)(k_z),
    \end{split}
\end{align}
for any fixed $k_y$ (for the demonstration, we set ${k_y = 0}$), where in the last inequality we used the positivity condition of the variance ({of $x$}) for the complexified HWFs. Hence, the QGB emerges, provided the evolution of $\bra{w^x_-} x \ket{w^x_-} = -\bra{w^x_+} x \ket{w^x_+} = \phi_x(k_z)/(2\pi)$, with $\ket{w^x_+}, \ket{w^x_-}$ realizing a $\mathcal{PT}$-symmetric pair of RTPs, which we further numerically observe in the bulk Wilson loops in App.~\ref{app::F}. Correspondingly, the spread of the maximally-localized HWFs needs to oscillate, as two opposite RTPs flow due to the non-trivial bulk real Hopf invariants. 

Finally, we remark that the quantum geometric breathing does not require the $C_2$ symmetry that is admitted by the considered models, as mentioned in the main text. Namely, as we retrieve numerically under the $C_2$ symmetry-breaking perturbations, the second-moment of the hybrid Wannier functions still correspondingly oscillates in the absence of $C_2$ symmetry, while the first moment remains vanishing in the real gauge under $\mathcal{PT}$ symmetry, as derived using the complexification trick. However, on breaking the $C_2$ symmetry, it is naturally possible for the maximally-localized hybrid Wannier functions to no longer respect the $C_2$ symmetry, despite the manifested presence of QGB, on evaluating their second moment. This shows that while the crystalline symmetries can effectively constrain the form and the functional evolution of QGB, the presence of QGB itself is not necessarily protected by these.

\section{Boundary states with surface Euler class}\label{app::F}

In this section, we analytically and numerically demonstrate the effective surface theories for real Hopf insulators (RHIs) with non-trivial strong invariants, as highlighted in the main text. 

We begin with an analytical derivation of the boundary theory for a three-band RHI. First, we modify the three-band Hamiltonian in Eq.~\eqref{eq:3BandFlat} of the main text as follows,
\begin{equation}
    H^{\text{cont}}_3(\kv) = 2 \textbf{d}(\kv) \otimes \textbf{d}(\kv)^\text{T} - |\textbf{d}|^2\mathbbm{1}_3,
\end{equation}
where we now use the unnormalized vector $\textbf{d} = z^\dagger \boldsymbol{\sigma}z$. This model realizes the same phase as the flattened Hamiltonian, as can be checked by explicitly computing the Hopf invariant of this phase. We now obtain an effective projected surface Hamiltonian
in the proximity of $z=0$. To do so, we first solve for the boundary eigenstates of $H^{\text{cont}}_3$, which take the form
\begin{equation}
      \ket{\psi_\pm} = \mathcal{N}_{+}
    \begin{pmatrix}
        0 \\
        1 \\
        \pm\ii 
    \end{pmatrix}
    e^{-(z \pm a)^2/2},  
\end{equation}
where $\mathcal{N}_{\pm}$ are normalization constants, and $a$ is a variational offset that paramaterizes the offset of these wavefunctions from the boundary at $z=0$. By projecting these states onto the bulk Hamiltonian $H^{\text{cont}}_3$ and taking the limit $a \rightarrow 0^\pm$ (which corresponds to taking the states to be infinitesimally separated from the boundary) one obtains an effective $2 \times 2$ Hamiltonian $H_{\text{eff}}^{s s'}=\bra{\psi_s} H^{\text{cont}}_3 \ket{\psi_{s'}}$, where $s, s'=\pm$, which manifestly respects the $\mathcal{C}_2\mathcal{T}$ symmetry at the boundary. The elements of this matrix are given by,
\beq{}
\begin{small}
    H_\text{eff} = 
    \begin{pmatrix}
    -[k_x^2 + k_y^2 + (a^2-1)]^2 & -4a^2 e^{-2 a^2} (k_y - \ii k_x)^2 \\
    -4a^2 e^{-2 a^2} (k_y + \ii k_x)^2 & - [k_x^2 + k_y^2 + (a^2-1)]^2 
    \end{pmatrix}
\end{small},
\eeq
which shows that far from the boundary $(a > 1)$, the Euler class deduced from $H_\text{eff}$ vanishes, unlike in the Bloch-Wannier states at the boundary ${(a \rightarrow 0)}$. In particular, the Euler class in $H_\text{eff}$ can be directly recognized by comparing to the topologically non-trivial Euler Hamiltonians of identical functional forms, as considered in Refs.~\cite{PhysRevX.9.021013,Morris_2024}. This concludes the analytical argument for the effective boundary theory of the three-band RHI, which deduces the presence of the surface Euler invariant.

For completeness, we also include an argument for the topology of the boundary surface states in the four-band RHIs, as obtained from a continuum effective surface theory. For the four-band case, we start by constructing a continuum bulk theory for the model $complex$ Hopf insulator $H_{\text{MRW}} = \vec{d}_{\hhc} \cdot \boldsymbol{\sigma} = (z^{\dagger} \boldsymbol{\sigma} z) \cdot \boldsymbol{\sigma}$~\cite{Hopf_1}, on expanding the vector $z$ to first order in momentum, $z = (k_x + \ii k_y, k_z + \ii m)^\text{T}$~\cite{alexandradinata2021}. In a matrix form, the continuum bulk Hamiltonian reads,
\beq{}
    H_{\text{MRW}} =
    \begin{pmatrix}
    k_x^2 + k_y^2 - m^2 - k_z^2 & ( k_x + \ii k_y)(k_z+ \ii m)\\
    ( k_x - \ii k_y)(k_z-\ii m) &  - k_x^2 - k_y^2 + m^2 + k_z^2.
    \end{pmatrix}
\eeq
\\
with Hopf invariant $\hhc = \frac{1}{2}\operatorname{sgn}(m)$.
In addition, we perform a substitution $k_z \rightarrow -\ii \partial_{z}$, obtaining:
\beq{}  H^{\text{cont}}_{\text{MRW}} =
    \begin{pmatrix}
     k_x^2 + k_y^2 - m^2 + \partial_{z}^2 & ( k_x + \ii k_y)(-\ii \partial_{z}+\ii m)\\
    ( k_x - \ii k_y)(\ii \partial_{z}-\ii m) &  - k_x^2 - k_y^2 + m^2 -  \partial_{z}^2
    \end{pmatrix}.
\eeq
We furthermore construct a parent $4 \times 4$ RHI Hamiltonian as $H^{\text{cont}}_4 = H^{\text{cont}}_{\text{MRW}} \oplus H^{\text{cont}}_{\text{MRW}}$, upon gluing two copies of complex Hopf insulators under $\mathcal{PT}$ symmetry. Correspondingly, we obtain an effective projected Hamiltonian with eigenvectors of $H^{\text{cont}}_4$ localized in the $z$-direction,
\begin{subequations}
\beq{}
    \ket{\psi_{+\pm}} = \mathcal{N}_{+\pm}
    \begin{pmatrix}
        1 \\
        0 \\
         1 \\
        0 \\
    \end{pmatrix}
    e^{-(z \pm a)^2/2},
\eeq
and for the other polarization in the initial orbital basis,
\beq{}
    \ket{\psi_{-\pm}} = \mathcal{N}_{-\pm}
    \begin{pmatrix}
        0 \\
        1  \\
        0 \\
        1 
    \end{pmatrix}
    e^{-(z \pm a)^2/2},
\eeq
\end{subequations}
where $\mathcal{N}_{+\pm}, \mathcal{N}_{-\pm}$ are normalization constants, and $a$ is again a variational offset constant from the boundary at $z=0$. Here, with the projection onto four states in the limit of infinitesimal proximity to the boundary $a \rightarrow 0$, one obtains the $4 \times 4$ Hamiltonian $H_{\text{b-ry}} = H'_{\text{eff}} \oplus H'_{\text{eff}}$, equivalent to two glued copies of Chern insulators under $\mathcal{C}_2\mathcal{T}$ symmetry enforced at the boundary. In terms of the aforementioned vectors, the $H_{\text{b-ry}}$ is more directly constructed as a matrix of elements $\bra{\psi_{\pm\pm}} H^{\text{cont}}_4 \ket{\psi_{\pm\pm}}$. Explicitly, the matrix is given by,
\beq{}
\begin{small}
    H'_\text{eff} = 
    \begin{pmatrix}
    k_x^2 + k_y^2 + (a^2-1) & a^2 e^{-2 a^2} (k_y - \ii k_x) \\
    a^2 e^{-2 a^2} (k_y + \ii k_x) & -k_x^2 - k_y^2 - (a^2-1) 
    \end{pmatrix}
\end{small},
\eeq
manifestly showing that far from the boundary $(a > 1)$, the Euler class deduced from $H'_\text{eff}$ vanishes, unlike in the Bloch-Wannier states at the boundary $(a \rightarrow 0)$, similarly to the three-band case. In particular, the Euler class in $H'_\text{eff}$ can be directly recognized, as in the three-band case. This concludes the analytical argument for the effective boundary theory of the four-band RHI, demonstrating the presence of the surface Euler invariant, and the relation to the continuum bulk-boundary physics of the three-band Hamiltonian. Finally, we note that to obtain the boundary spectrum of a three-band RHI from a four-band RHI, alternatively, the limit of $E_4 \rightarrow \infty$ could be taken, which trivializes one of the subblocks of the $H'_{\text{eff}}$, while keeping the surface Euler invariant in the other block (corresponding to the occupied states) intact. 

\begin{figure*}
\centering
  \includegraphics[width=1.0\linewidth]{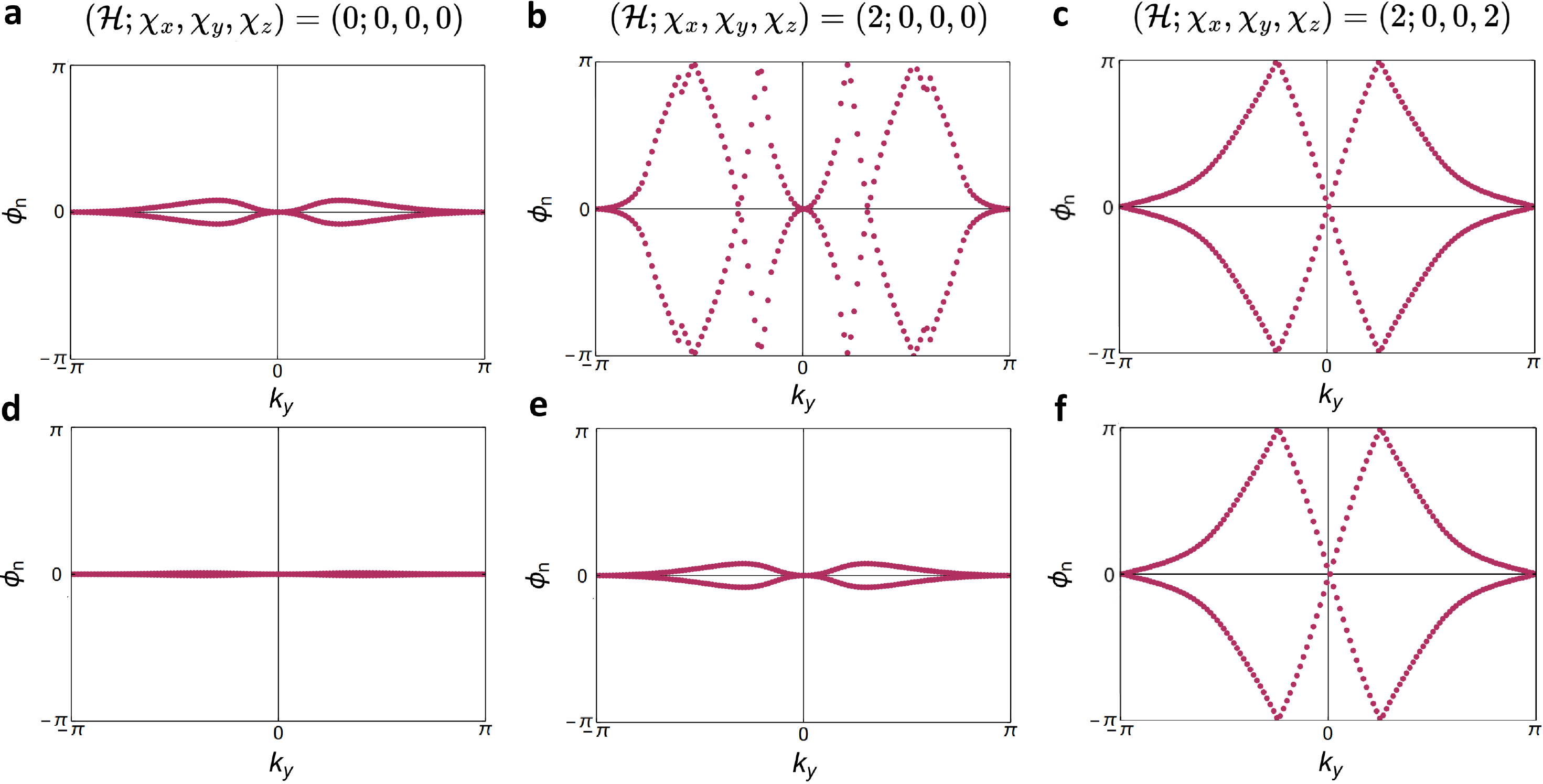}
  \caption{The interplay of bulk and surface invariants in three-band real Hopf insulators; bulk. (\textbf{a}--\textbf{c}) Bulk Wilson loops for different values of the Hopf-Euler invariants $(\mathcal{H}; \chi_x, \chi_y, \chi_z)=(0;0,0,0), (2;0,0,0), (2;0,0,2)$ at $k_z = 0$, In the last Hopf-Euler case, the Wilson loop windings corresponds to the Wilson loop winding of the Euler phases before extending/pumping.
  (\textbf{d}--\textbf{f}) Bulk Wilson loops for different values of the Hopf-Euler invariants ${(\mathcal{H}; \chi_x, \chi_y, \chi_z)=(0;0,0,0), (2;0,0,0), (2;0,0,2)}$ at $k_z = \pi$, i.e. mid-way through the cycle. We numerically find that ${2\hh  = \#(k_z = 0) + \#(k_z = \pi})$, where $\#$ denotes the number of crossings at $\phi_n = \pi$, corresponding to the flow of the pairs of opposite RTPs. The eigenvalues corresponding to the opposite RTPs touch at $k_z=0$, rather than cross; unlike in the case of the non-trivial weak Euler invariants, where a crossing occurs (\textbf{c},\textbf{f}). We conclude that to realize an Euler invariant $\chi_z$ in a default configuration ($k_z = 0$) within the Hopf-Euler insulator, the Hamiltonian with $\hh  = \chi$ may be chosen, but nonetheless the presence of the weak invariant is necessary.}
\label{figWilson}
\end{figure*}

Furthermore, we numerically validate the argument about the topologies of the bulk states and boundaries realized in the proposed three-dimensional three-band Hamiltonians. First, we show the bulk Wilson loops of the three-band RHI, which, in particular, reflect the presence of both strong and weak invariants; see Fig.~\ref{figWilson}. Having diagnozed topologically non-trivial bulk, as shown in Fig.~\ref{figWilson}, we next study the topological character of the boundary. Namely, we first show the Wannier-Stark ladders of the RHIs realizing the strong Hopf invariant, which similarly to Fig.~\ref{figWilson}, explicitly demonstrates the complexified returning Thouless pump (RTP); see also App.~\ref{app::E} for more general details.  Moreover, from the surface Wilson loop winding, we observe that manifestly, $\chi_s = \hh$. 
Numerically, to construct the Wannier ladders, the hybrid Wannier functions (HWFs), maximally localized in the $y$-direction, were evaluated as the eigenfunctions of the position operator $y$-component projected on the occupied two-band subspace. The projector on ground state $\hat{P}=\sum^{\text{occ}}_n \ket{\psi_n}\bra{\psi_n}$ was constructed from the occupied energy eigenstates as the Hamiltonian was Fourier transformed in the $y$ direction and a chain with 20 sites under open boundary conditions was considered. The operator $\hat{P}\hat{y}\hat{P}$ was diagonalized on a mesh in the 2D reduced Brillouin zone of quasimomenta ($k_x$, $k_z$). The HWFs and their centers ($\bar{w}_y$) in the $y$-direction were directly extracted as the eigenvectors and the eigenvalues of the problem.\\
\begin{figure*}
\centering
  \includegraphics[width=1.0\linewidth]{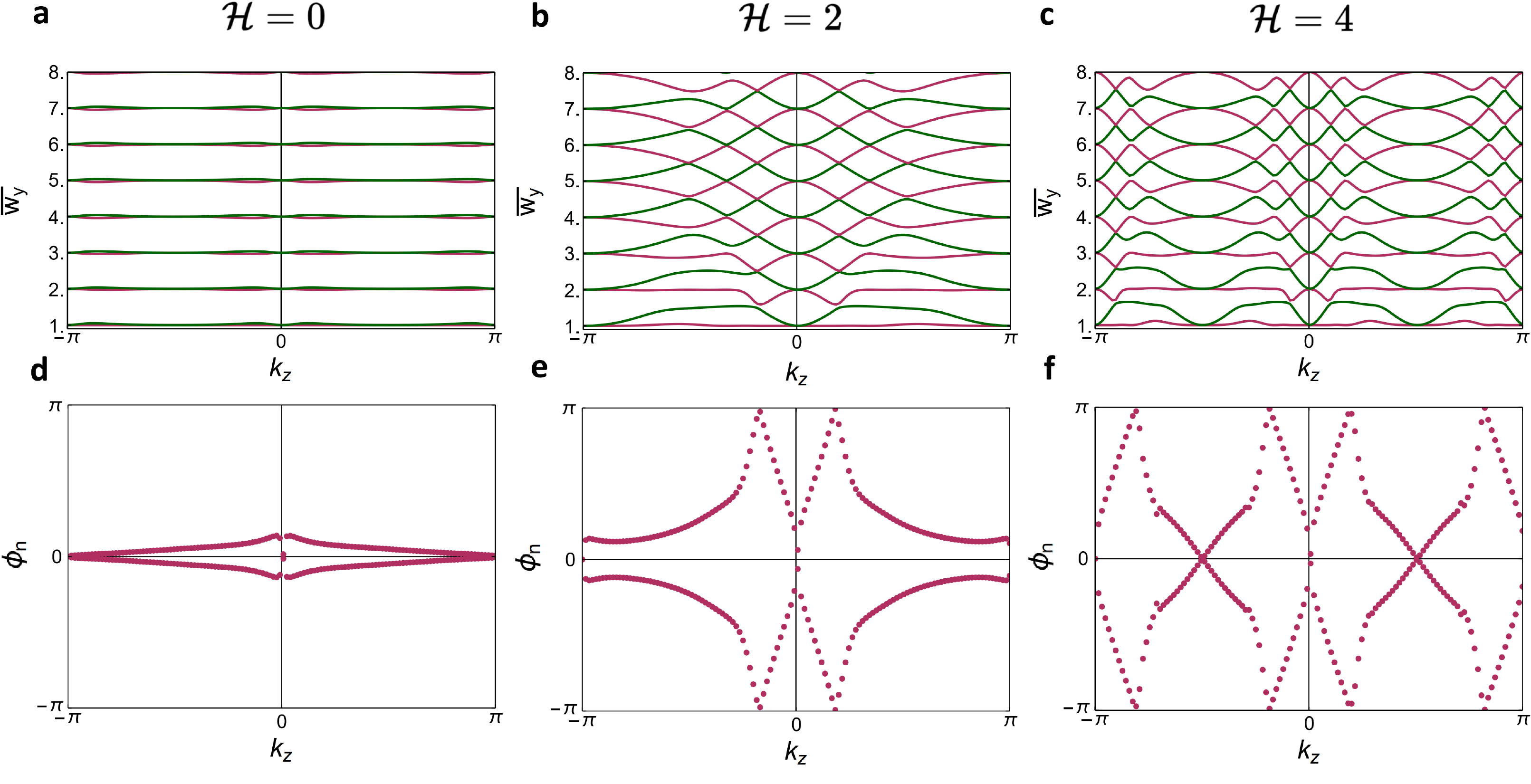}
  \caption{The interplay of bulk and surface invariants in three-band real Hopf insulators; surface. (\textbf{a}--\textbf{c}) Wannier-Stark ladders, for different values of the Hopf invariant $\hh =0,2,4$, with the bottom two Bloch-Wannier bands at the boundary \textit{separated} from the bulk with a $\mathcal{C}_2\mathcal{T}$-preserving perturbation.
  (\textbf{d}--\textbf{f}) Surface Wilson loop windings for bulk Hamiltonians with $\hh =0,2,4$, demonstrating the surface Euler numbers $\chi_s = \hh $ manifested by the three-band RHIs, and hence by the three-band strong Hopf-Euler phases.}
\label{figBBC}
\end{figure*}
Additionally, to compute the surface Wilson loop winding, the surface Bloch-Wannier bands were \textit{isolated} from the bulk by an inclusion of a $\mathcal{C}_2\mathcal{T}$-preserving perturbation ($V_{\mathcal{C}_2\mathcal{T}}$) to the projected Hamiltonian,
\beq{}
    V_{\mathcal{C}_2\mathcal{T}} =
    \begin{pmatrix}
    -0.4 & 0 & 0 \\
    0 & -0.1 & 0 \\
    0 & 0 & 0.2
    \end{pmatrix},
\eeq
achieving a separation of surface Wannier bands from the bulk bands on both the opposite facets of the real Hopf insulators under open boundary conditions in $y$-direction.  

Having isolated the surface Bloch-Wannier bands from the bulk, a standard procedure of computing non-Abelian Wilson loops was employed. Here, the winding was explicitly retrieved from the surface bands given by $\ket{u^s_{n}(k_x,k_y)}$ (see also Fig.~\ref{figBBC}), and as predicted from an analytical argument, the Wilson loop eigenvalues $(\phi_n)$ indicate the presence of the surface Euler numbers ($\chi_{s}$). The surface Euler numbers are defined as,
\beq{}
    \chi_s = \frac{1}{2\pi}\int_{\text{rBZ}} \dd^2\kv~ \text{Eu}_{xy,s}, 
\eeq
with the integration performed here over the reduced Brillouin zone, ${\text{rBZ} = \{(k_x, k_y)\} \cong T^2}$, and the integrand, the surface Euler curvature ${\text{Eu}_{xy,s} \equiv \bra{\partial_{k_x} u^s_{[1}(k_x,k_y)}\ket{\partial_{k_y} u^s_{2]}(k_x,k_y)}}$, with $[\ldots]$ denoting the antisymmetrization with respect to the band indices. Our result is also consistent with the finding of the presence of opposite surface Chern numbers ($C_s$) on the boundaries of the complex two-band Hopf insulators~\cite{alexandradinata2021}, which is supported by the complexification correspondence, on breaking the $\mathcal{C}_2\mathcal{T}$ symmetry~\cite{bouhon2022multigap}. For completeness, we reiterate that in the Hopf insulators, the $C_s$ invariant reads 
\beq{}
    C_s = \frac{1}{2\pi} \int_{\text{rBZ}} \dd^2\kv~ \Omega^n_{xy,s}, 
\eeq
with $\Omega^n_{xy,s} \equiv \ii[\bra{\partial_{k_x} u^s_{n}(k_x,k_y)}\ket{\partial_{k_y} u^s_{n}(k_x,k_y)} - \text{c.c.} ]$ the surface Berry curvature, in terms of the Bloch-Wannier states $\ket{u^s_{n}(k_x,k_y)}$. However, distinctively, the surface Euler invariant -- as derived from the real Hopf invariant ($\hh$) inducing an effective surface Euler theory within our models -- was obtained in three bands, under $2 \oplus 1$ partitioning, which is different from the other reported Hopf phases~\cite{alexandradinata2021,Zhu_2023,Lim2023}.

 \section{Homotopy invariants and complexification correspondences}\label{app::G}
 \begin{figure*}
 \centering
   \includegraphics[width=1.5\columnwidth]{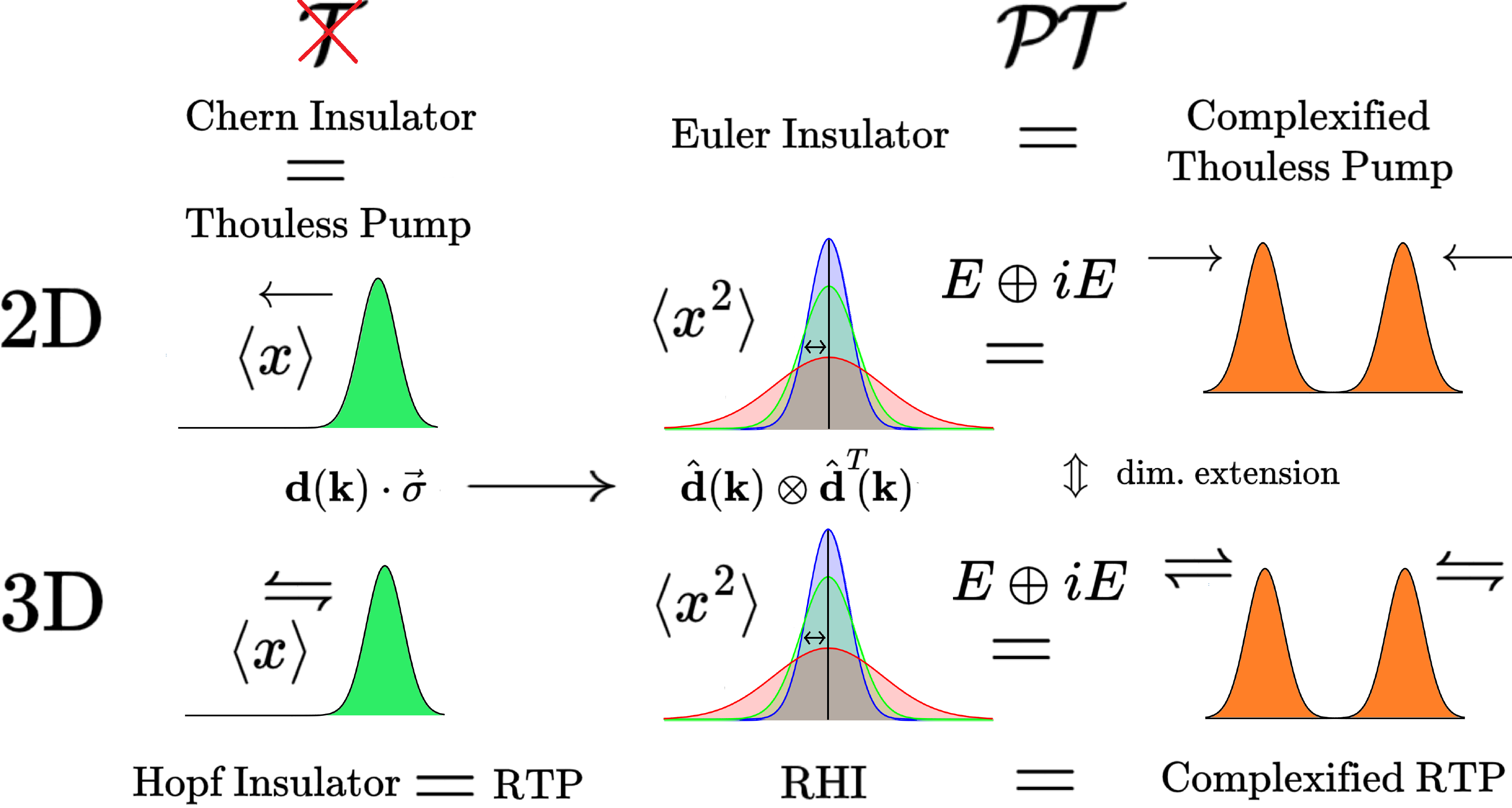}
   \caption{The family of homotopy-classified topological phases and relations between them, demonstrated in terms of hybrid Wannier functions (HWFs) and their distinct evolutions defining a class of Thouless pumps~\cite{PhysRevB.27.6083}. The complexification of the RTP of the Hopf insulator is realized in the three-band RHI, yielding QGB (see also App.~\ref{app::E}). As the QGB occurs with $k_z$ changing, the HWFs oscillate in $x$ and $y$, as their second moment $\langle x^2 \rangle$ changes, rather than flowing (which would be equivalent to {$\langle x \rangle$ changing}), as in the Thouless pump, where quantized charge is pumped. The second moment is reflected by the optical, rather than DC transport properties, which are governed by the first moment. In the case of three-band Hopf phases, the topological invariant is optically manifested through the quantized shift response.}
\label{figPump}
\end{figure*}

In this section we discuss the relations between the various topological invariants that characterize two-band complex phases, and three-band real phases, in two and three dimensions (see Fig.~\ref{figPump}). In particular, we relate the complex Chern and Hopf invariants to the real Euler and Hopf invariants that exist in the multi-gap topological phases considered in this work.

We begin with two-band complex phases. The Hamiltonians of such systems may always be written in the form $H(\vb{k})=\vb{d}(\vb{k})\cdot\boldsymbol{\sigma}$ (ignoring contributions $\propto\mathbbm{1}_2$), where $\vb{d}(\vb{k})$ is a three component real vector that, when normalized, defines a map $\dh:\BZ\to S^2$. The two-sphere arises here as it is the classifying space of these systems, equal to the Grassmannian
\begin{equation}
    \mathrm{Gr}_{1, 2}(\mathbb{C}) = \frac{\mathsf{U}(2)}{\mathsf{U}(1)\times\mathsf{U}(1)}\cong S^2.
\end{equation}

In two dimensions, one can compute the Chern number $C$ of the occupied subspace of a complex Hamiltonian from the integral of the Berry curvature $F=\dd A$ over the Brillouin zone, where $A=\ii\braket{u}{\dd u}$ is the Berry connection of the occupied band $\ket{u}$. Note that, for the Chern number to be non-vanishing, the system must violate time-reversal symmetry $\mathcal{T}$. In the particular case of a two band system, $C$ is an element of the homotopy group $\pi_2[S^2] \cong \mathbb{Z}$, and is equal to the `wrapping number' of the map $\dh(\vb{k})$:
\begin{equation}\label{eq:ChernNum}
    C = \frac{1}{2\pi}\intBZ F=\frac{1}{4\pi} \int \dd^2\kv~ \hat{\textbf{d}} \cdot (\partial_{x} \hat{\textbf{d}} \cross \partial_{y} \hat{\textbf{d}}).
\end{equation}
where  $\partial_i \equiv \partial_{k_i}$. The integrand in the second part of Eq.~\eqref{eq:ChernNum} may be interpreted as the skyrmion density of the vector field defined by $\hat{\textbf{d}}$. 

In three dimensions, the vector $\dh$ defines a map $T^3\sim S^3\to S^2$, so in this case the system can be assigned a topological index in the homotopy group $\pi_3[S^2] \cong \mathbb{Z}$. This index is known as the Hopf invariant, and it may be computed from the integral of the the Abelian Chern-Simons form $A\wedge\dd A=A\wedge F$ over the BZ~\cite{Hopf_1}, 
\begin{align}\label{eq:ComplexHopf}
\begin{split}
    \hhc ={}& -\frac{1}{4\pi^2}\intBZ A\wedge F \\
    ={}&-\frac{1}{4\pi^2}\intBZ \dd^3\vb{k}~ \varepsilon_{ijk} \hat{\vb{Z}}^{\dagger}(\partial_{i} \hat{\vb{Z}}) (\partial_j \hat{\vb{Z}}^{\dagger})(\partial_{k} \hat{\vb{Z}}).
\end{split}
\end{align}
In the second line of Eq. \eqref{eq:ComplexHopf} we have given an alternative expression for $\hhc$ involving a (normalized) two component complex vector $\vb{Z}=\vb{Z}(\vb{k})$, in terms of which the winding vector is written as $\hat{d}_i = \vb{Z}^\dagger\sigma_i\vb{Z}$. As discussed extensively in the main text, the Hopf invariant has a geometric interpretation as the linking number of the preimages of two arbitrarily chosen points on $S^2$ under $\dh$. Alternatively, one may interpret the texture of $\dh(\vb{k})$ as a `Hopfion' in momentum space, similarly to the two-dimensional Skyrmion which realizes a Chern number.

In addition to the Hopf index, one may compute a separate Chern number on each of each of two dimensional coordinate planes inside the BZ. These are computed by applying Eq. \eqref{eq:ChernNum} to the restricted maps $\hat{\vb{v}}_i=\dh(\vb{k})|_{k_i=\text{const.}}$, $i=x, y, z$. Provided the valence and conduction bands are gapped at all points in the BZ, these Chern numbers are independent of the particular value of $k_i$ chosen.

We now discuss the topological invariants for real, three-band systems. Due to the reality condition, the classifying space of these Hamiltonians is now a quotient of orthogonal, rather than unitary groups, and for a system with two occupied bands it is given by
\begin{equation}
    \mathsf{Gr}_{2, 3}(\mathbb{R})= \frac{\mathsf{O}(3)}{\mathsf{O}(2)\times\mathsf{O}(1)}\cong \mathbb{R}P^2.
\end{equation}
In contrast to two-band Hamiltonians, the map $T^3\to\mathbb{R}P^2$ is not given by a vector appearing in the parametrization of the Hamiltonian, but instead by the third eigenvector $\ket{u_3(\vb{k})}$. By flattening the bands of the Hamiltonian we can always write 
\begin{equation}
    \bar{H}_3(\vb{k}) = 2\vb{u}_3(\vb{k})\otimes\vb{u}_3(\vb{k}) -\mathbbm{1}_3,
\end{equation}
which gives an explicit expression for the flattened Hamiltonian in terms of this map. 

Since the real projective plane $\mathbb{R}P^2$ is isomorphic to $S^2/\mathbb{Z}_2$, many of the topological properties of two-band complex Hamiltonians have analogues in three-band real Hamiltonians. Firstly, in two dimensions the topological invariant corresponding to the homotopy group $\pi_2[\mathbb{R}P^2]\cong\pi_2[S^2]\cong\mathbb{Z}$ -- the wrapping number of the sphere -- is the Euler class~\cite{Bouhon_2019},
\begin{equation}\label{eq:EulerSkryme}
    \chi=\frac{1}{2\pi}\intBZ \Eu = \frac{1}{2\pi}\int\dd^2\vb{k}\,\vb{u}_3\cdot(\partial_x\vb{u}_3\cross\partial_y \vb{u}_3).
\end{equation}
The above expression for the Euler class in terms of the skyrmion density third eigenvector is special to three-band systems and is a consequence of the relation $\vb{u}_3=\vb{u}_1\cross\vb{u}_2$. In general, the Euler class is a multi-band invariant that characterizes the topology of the occupied two-band subspace spanned by two bands $\vb{u}_{1, 2}$. It is calculated using the Euler form ${\Eu =\dd\aa}$, where $\aa=\Pf[-\ii \braket{u_i}{\dd u_j}]$ is the Euler connection, which is equal to the Pfaffian of the non-Abelian Berry connection in the occupied subspace.

Aside from the obvious similarity of Eqs.~\eqref{eq:ChernNum} and \eqref{eq:EulerSkryme}, the Chern number and the Euler class are related through complexification, in the sense that the Euler class of a subspace spanned by the real eigenvectors $\ket{u_{1, 2}}$ is equal to the Chern number of the single complex state $\ket{v}=(\ket{u_1}+\ii\ket{u_2})/\sqrt{2}$, i.e. $\chi\left[\ket{u_1}, \ket{u_2}\right] = C[\ket{v}]$. This formula is useful for deriving properties of the multi-band Euler class from those of the single-band Chern class (see~App.~\ref{app::E}).

As a central component of our work, we recognize that in three dimensions, the eigenvector of $H_3$ defines a map $\vb{u}_3: T^3\sim S^3\to S^2$, meaning that real three-band phases in 3D may display a Hopf index ${\hh\in\pi_3[\mathbb{R}P^2]\cong\pi_3[S^2]\cong \mathbb{Z}}$. Writing $(u_3)_i=\vb{z}^\dagger\sigma_i\vb{z}$, where $\vb{z}(\vb{k})$ is, like $\vb{Z}(\vb{k})$ in Eq. \eqref{eq:ComplexHopf}, a two-component complex vector of unit norm, we can express the Hopf index as
\begin{align}
\begin{split}
    \hh ={}& -\frac{1}{16\pi^2}\intBZ \aa\wedge \Eu \\
    ={}&-\frac{1}{4\pi^2}\intBZ \dd^3\vb{k}~ \varepsilon_{ijk} \hat{\vb{z}}^{\dagger}(\partial_{i} \hat{\vb{z}}) (\partial_j \hat{\vb{z}}^{\dagger})(\partial_{k} \hat{\vb{z}}).
\end{split}
\end{align}
When $\hh\neq 0$, the third eigenvector displays a nontrivial Hopfion texture in momentum space like that described for the winding vector $\dh$ above. 

Finally, we note that in three spatial dimensions, the system may also host an Euler class on each of the coordinate planes within the BZ. The subdimensional Chern numbers discussed above are computed with Eq. \eqref{eq:ChernNum}, and similarly these Euler classes are computed with Eq. \eqref{eq:EulerSkryme}. Moreover, they are independent of the particular coordinate slice used to compute them. The interplay of the real Hopf index and the subdimensional Euler classes is a central focus of this work.

\end{document}